\documentstyle [12pt,epsfig]{article} 
\makeatletter
\@addtoreset{equation}{section}
\makeatother
\renewcommand{\theequation}{\thesection.\arabic{equation}}
\newcommand{\ba}{\begin{eqnarray}}
\newcommand{\ea}{\end{eqnarray}}

\newcommand{\E}{{\cal E}}
\newcommand{\B}{{\cal B}}
\newcommand{\A}{{\cal A}}
\newcommand{\Ei}{{\rm E}}
\newcommand{\Bi}{{\rm B}}
\newcommand{\Ai}{{\rm A}}
\newcommand{\Ji}{{\rm J}}
\newcommand{\Mg}{{\cal M}_{\gamma}}

\begin{document}
\newcommand{\BS}{\bigskip}
\newcommand{\SECTION}[1]{\BS{\large\section{\bf #1}}}
\newcommand{\SUBSECTION}[1]{\BS{\large\subsection{\bf #1}}}
\newcommand{\SUBSUBSECTION}[1]{\BS{\large\subsubsection{\bf #1}}}

\begin{titlepage}
\begin{center}
\vspace*{2cm}
{\large \bf Space-time transformation properties 
   of inter-charge forces and dipole radiation: Breakdown of the classical field concept in relativistic electrodynamics}
\vspace*{1.5cm}
\end{center}
\begin{center}
{\bf J.H.Field }
\end{center}
\begin{center}
{ 
D\'{e}partement de Physique Nucl\'{e}aire et Corpusculaire
 Universit\'{e} de Gen\`{e}ve . 24, quai Ernest-Ansermet
 CH-1211 Gen\`{e}ve 4.}
\end{center}
\begin{center}
{e-mail; john.field@cern.ch}
\end{center}
\vspace*{2cm}
\begin{abstract}
   A detailed study is made of the space-time transformation properties of intercharge forces
  and the associated electric and magnetic force fields, both in classical electromagnetism and
  in a recently developed relativistic classical electrodynamical theory. Important differences
  are found and serious errors are found in the traditional treatment of special-relativistic
  effects in classical electromagnetism. Fields associated with radiation processes are also considered
   and classical
   and quantum mechanical predictions are compared. An important, though trivial, mathematical
   error in the derivation leading to the retarded Li\'{e}nard-Wiechert retarded potentials is
   pointed out. It is also shown why electric and magnetic fields produced by moving
   source charges or current distributions are not `classical fields', and that they do not,
   in general, constitute an antisymmetic second-rank tensor, as is usually assumed.

\end{abstract}
\vspace*{1cm}{\it Keywords};  Special Relativity, Classical Electrodynamics.
\newline
\vspace*{1cm}
 PACS 03.30+p 03.50.De
\end{titlepage}

\SECTION{\bf{Introduction}}
 In a recent paper by the present author ~\cite{JHF1}, the mechanical aspects of classical 
 electrodynamics, that is the description of the forces between electric charges with arbitary
  relative motion, was developed from the three postulates:
  \begin{itemize}
   \item[(i)] Coulomb's Law
   \item[(ii)] Special relativistic invariance
 \item[(iii)] Hamilton's Principle
 \end{itemize}
  The forces between two charges of magnitude $q_1$, $q_2$ are derived from the manifestly
  Lorentz invariant Lagrangian:
  \begin{equation}
   L(x_1,u_1;x_2,u_2) = -\frac{m_1 u_1^2}{2} -\frac{m_2 u_2^2}{2} - \frac{j_1 \cdot j_2}{c^2 \sqrt{-(x_1-x_2)^2}}
  \end{equation}
   where $c$ is the speed of light in vacuum.
   Here the space-time positions of the charges are specified by the 4-vectors $x_1$, $x_2$. $u_1$ and $u_2$ are
   the 4-vector velocities of the charges and $j_1$ and $j_2$ their 4-vector electromagnetic currents, $q_1 u_1$
 and $q_2 u_2$, respectively. In accordance with quantum electrodynamics~\cite{JHF1}, the interaction between the
  charges is instantaneous\footnote{A time-like metric is employed for 4-vector products}:
   \[   t_1 = t_2,~~~~ -(x_1-x_2)^2 = (\vec{x}_1 -\vec{x}_2)^2 = r^2 \]
    in the overall centre-of-mass system of the interacting charges. 
  \par The non-retarded nature of electrodynamical force fields and
    their consistency with instantaneous
   action-at-a-distance has recently been demonstrated in an experiment measuring the delay of a signal
   induced by electromagnetic induction as a function of the separation of the transmitting and receiving
   antennae~\cite{JAP1}
  \par Inserting the Lagrangian (1.1) into the covariant Lagrange equations yields the first-order differential
  equations that describe the motion of the charges:
    \begin{eqnarray} 
      \frac{d\vec{p_1}}{dt} &  = & \frac{q_1}{c}\left[\frac{ j_2^0\vec{r} +  \vec{\beta}_1 \times
     (\vec{j_2} \times \vec{r})}{r^3} -\frac{1}{c r}\frac{d \vec{j_2}}{d t}-\vec{j_2}
     \frac{(\vec{r} \cdot \vec{\beta}_2)}{r^3} 
      \right] \\
       \frac{d\vec{p_2}}{dt}  &  = & -\frac{q_2}{c}\left[\frac{ j_1^0\vec{r} +\vec{\beta}_2 \times 
  (\vec{j_1} \times \vec{r})}{r^3}+\frac{1}{c r}\frac{d \vec{j_1
}}{d t} -\vec{j_1}
     \frac{(\vec{r} \cdot \vec{\beta}_1)}{r^3}
    \right] 
     \end{eqnarray} 
  where $\vec{p}_1 = \gamma_1 m_1 \vec{v}_1$,  $\vec{p}_2 = \gamma_2 m_1 \vec{v}_2$, $\vec{v}_1$,  $\vec{v}_2$
   are the 3-vector velocities of the charges and $\beta$, $\gamma$ are the usual dimensionless parameters of special
    relativity:
   \[ \beta \equiv \frac{v}{c},~~~\gamma \equiv \frac{1}{\sqrt{1-\beta^2}} \]
    Equations (1.2) and (1.3) provide a complete description of the classical relativistic dynamics of
    two isolated charges undergoing mutual electromagnetic interaction. They contain however no `fields'
     and the force-field concept was not used in their derivation. It is evidently of interest, however, to compare 
    the forces given by the right sides of (1.2) and (1.3) with those generated by the force fields of 
    conventional classical electromagnetism, referred to in the following, for brevity, as CEM. The latter is 
    described in detail in many text books, for example, the well-known ones of Landau and Lifshitz~\cite{LL}
    and Jackson~\cite{Jack1}. The force fields with $1/r^2$ dependence are associated, in QED, with the exchange
    of space-like virtual photons~\cite{JHF1}, whereas classical radiation fields with
     $1/r$ dependence, to be dicsussed in Section 5 below, correspond to the QED processes of production
     and propagation of real photons moving at speed $c$. The first four sections of the present paper are then mainly devoted to a detailed
     comparison of
    the predictions of (1.2) and (1.3), referred to below as RCED, for `Relativistic Classical Electro-Dynamics',
    with the conventional CEM predictions. It is pointed out that, contrary to the usual assumption,
     neither the 4-vector potential, nor electric and magnetic fields are `classical fields' in the sense
     explained in Section 3. In Section 4 it is demonstrated that the longitudinal electric field of 
    a uniformly moving charge is not covariant, and that CEM fields do not obey the electric 
    field divergence equation or the electrodynamic Maxwell equation\footnote{The `electrodynamic'
    Maxwell equation, is the generalised Amp\`{e}re law that contains the time derivaive of the 
    electric field (Maxwell's `displacement current', the physical origin of which
     is discussed in detail in Section 4 below). The `magnetodynamic'  Maxwell equation, containing
    the time derivative of the magnetic field, is the Faraday-Lenz law.} (Amp\`{e}re's law). A corollary
    is that, as previously pointed out, the electric field of a moving charge does not obey
     Gauss' law~\cite{JHF2}.
    \par It has previously been remarked that, when terms of O($\beta^2$) and higher are taken into account,
     the predictions
    of CEM and RCED are not the same. In particular, it has been shown that the electric field of a uniformly
     moving charge given by the Heaviside formula~\cite{Heaviside}, gives inconsistent results for Rutherford 
    scattering in different inertial frames~\cite{JHF2}. The fields of an accelerated charge derived from `retarded'
     Li\'{e}nard Wiechert (LW) potentials~\cite{LW}, as used in the derivation of the Heaviside formula, are
     found to be,
     unlike the RCED forces, incompatible with the existence of stable, circular, Keplerian orbits~\cite{JHF2}. 
     Jackson's `Torque Paradox'~\cite{JackTP} where application of a Lorentz transformation apparently induces a 
     torque on two equal, mutually interacting, electric charges, and which is a consequence of the use of the
     Heaviside formula and the associated magnetic field, is resolved~\cite{JHF2} by the use of (1.2) and (1.3).
     No torque is produced by the RCED forces.  It has also been demonstrated that use of the
      Heaviside formula predicts the absence of electromagnetic induction in a frame where a magnet
    is in motion and the test charge is at rest, for an elementary `magnet'
    consisting of two moving charges in a symmetric configuation~\cite{JHF3}.
    In contrast, consistent results are found both in the frame where
    the `magnet' is in motion, and in the one where it is at rest, by use of (1.2) and (1.3).
    \par In Section 5 radiation fields are considered. They are obtained, in the conventional manner, as solutions
    of inhomogeneous d'Alembert equations derived by combining the electrodynamic Maxwell
    Equation with the Lorenz condition\footnote{Not `Lorentz condition', as found in many text books~\cite{JackOkun}.} for the 4-vector potential. The retarded solutions proportional 
    to $1/r^2$ are found to differ from the conventional ones derived from the LW
    potentials, but to agree, up to a common factor and a retarded time argument, with the RCED force
   fields corresponding to (1.2) and (1.3). The difference is shown to result from an incorrect treatment
    of the effective density of a moving charge distribution in deriving the LW potentials. Thus the standard
    CEM force fields are incorrect, firstly because they are assumed to be retarded rather than instantaneous
    and secondly because there is a mathematical error in their derivation. The retarded solutions of the
   d'Alembert equations, proportional to $1/r$, describing radiative effects (the creation and propagation
   of real photons) are discussed for the case of dipole radiation in the far zone and shown to be
    equivalent, in the non-relativistic limit and for low photon densities, with quantum mechanical predictions
     for a static dipole source. However radiative effects in a boosted frame with $\beta \simeq 1$ are 
    found to be not correctly described by the Poynting vector of the transformed radiation fields in this frame.
    \par Section 6 contains a summary and outlook. Because of the length of the paper, the reader is encouraged
     to read this section first, in order to obtain an overview of the material presented, and to return
     later to the previous sections for further information and derivations. Details of the more
    lengthy calculations are deferred to three appendices.

\SECTION{\bf{Space-time transformation properties of inter-charge forces in RCED}}\

\begin{figure}[htbp]
\begin{center}
\hspace*{-0.5cm}\mbox{
\epsfysize15.0cm\epsffile{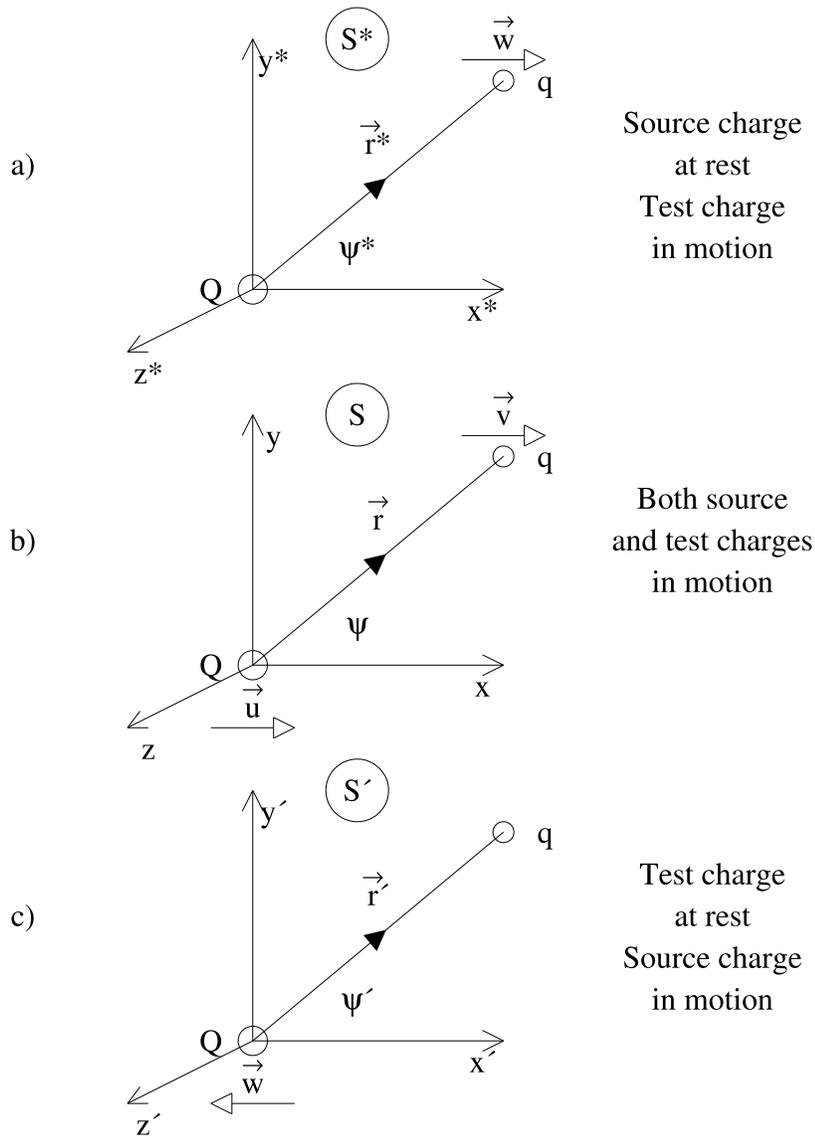}}   
\caption{{\sl Reference frames used to discuss the transformation laws of intercharge forces and
  force fields. Different polar angles and charge separations are indicated in each frame.
   However, it is demonstrated in Section 3 that $|\vec{r}^*| = |\vec{r}| = |\vec{r}'|$ and
    $\psi^* = \psi = \psi'$, as a consequence of the Lorentz invariant
    nature of the intercharge separation.}}
\label{fig-fig1}
\end{center}
\end{figure}
 
  In the following, one charge of magnitude $Q$ will be considered as the `source' of the force $\vec{F}$ that is
   exerted on the `test charge' of magnitude $q$. In this section the relation between the values of $\vec{F}$
   found in different frames will be analysed. For this, it will be convenient to introduce the three
   inertial frames shown in Fig.1. In the frame S$^*$ (Fig.1a) the source charge is at rest and the test charge
   moves with speed $w$ parallel to the +ve x-axis. In the frame S (Fig.1b) both charges are in motion parallel to
   the +ve x-axis, $Q$ with speed $u$, $q$ with speed $v$. Finally in S'(Fig.1c), the test charge is at rest
   while the source charge moves with speed $w$ along the -ve x-axis. With the definitions:
   \[ \beta_u \equiv \frac{u}{c},~~~\gamma_u \equiv \frac{1}{\sqrt{1-\beta_u^2}} \]    
   $w$, $u$ and $v$ are related by the parallel velocity addition formula:
   \begin{equation}
    \beta_w = \frac{\beta_v-\beta_u}{1-\beta_v\beta_u}
    \end{equation}
     Introducing unit vectors $\hat{\imath}$, $\hat{\jmath}$, $\hat{k}$ along the x-, y- and z-axis directions
  and the angle $\psi$ between the radius vector joining the source and test charges and the +ve x-axis, 
   the force $\vec{F} = d\vec{p}/dt$ on the test charge in the different frames shown in Fig.1 is given by Eqn(1.2)
   as\footnote{ Since the source charge is in uniform motion, the term $\simeq d \vec{j}_2/dt$ in (1.2) vanishes.}:
   \par \underline{S$^*$, source charge at rest}
   \begin{equation}
    \vec{F}^* = \frac{q Q}{r^2}[\hat{\imath}\cos \psi + \hat{\jmath}\sin \psi] = \frac{q Q \vec{r}}{r^3}
   \end{equation}
    \par \underline{S, source and test charge in motion}
   \begin{equation}
    \vec{F} = \frac{q Q}{r^2}\left[\frac{\hat{\imath}\cos \psi}{\gamma_u} + \hat{\jmath} \gamma_u(1-\beta_u
     \beta_v)\sin \psi\right]
   \end{equation}
  \par \underline{S', test charge at rest}
  \begin{equation}
    \vec{F}' = \frac{q Q}{r^2}\left[\frac{\hat{\imath}\cos \psi}{\gamma_w} +
   \hat{\jmath} \gamma_w \sin \psi\right]
   \end{equation}
    In these equations the Lorentz invariant character of the spatial separation of the charges, mentioned
    above, which is also derived, from first principles,\footnote{That is, from the definition of the spatial
    separation of the charges at some instant in an arbitary reference frame.} from special relativity, in
      Section 3 below, is taken
    into account. Thus $r^* = r = r'$ and $\psi^*= \psi = \psi'$. 
   In all cases the force vector lies in the x-y plane, but it is radial only in rest frame of 
   the source charge. As shown in Ref.~\cite{JHF2}, the non-radial behaviour of the force in S' implies a similar
    behaviour of the corresponding electric field which therefore does not obey Gauss' theorem. 
    \par The transformation laws of the x- and y-components of the forces may be read off from 
     Eqns(2.2)-(2.4) with the results:
    \begin{eqnarray}  
     F_x & = & \frac{1}{\gamma_u} F_x^*,~~~  F_y = \gamma_u(1-\beta_u \beta_v) F_y^*
      = \frac{\gamma_w}{\gamma_v} F_y^* \\
    F'_x & = &\frac{1}{\gamma_w} F_x^*,~~~  F'_y = \gamma_w F_y^* \\
   F_x & = &\frac{\gamma_w}{\gamma_u} F'_x,~~~  F_y = \frac{1}{\gamma_v} F'_y 
     \end{eqnarray}
  In (2.5) the relation $\gamma_w = \gamma_u(\gamma_v-\beta_u \beta_v \gamma_v)$ which is the Lorentz
   transformation of the temporal component of the dimensionless 4-vector velocity ($\gamma_v$;$\gamma_v \beta_v$,0,0) has
   been used. Inspection of (2.5) resolves immediately Jackson's `Torque Paradox'~\cite{JackTP}. In this case 
   a comparison is made of the configuration in Fig.1a when $w = 0$ with that in Fig.1b when $v = u$. 
    Eqn(2.2) shows that the force is radial in S*. If  $w = 0$  and  $v = u$ Eqn(2.5) gives $F_x = F_x^*/\gamma_u$,
    $F_y = F_y^*/\gamma_u$ so that the force is also radial in S ---there is no torque in this frame.

\SECTION{\bf{Space-time transformation properties of electric and magnetic force-fields in RCED and CEM}}
 `Electric' and `magnetic' force-fields may be defined in RCED by comparing Eqn(1.2) with  the Lorentz force law
 \footnote{In this way, as shown in Ref. ~\cite{JHF1}, both the Lorentz force law and the definitions 
  of $\vec{E}(RCED)$ and $\vec{B}(RCED)$, in terms of the charge currents and the intercharge
   separation, are obtained from the `fieldless' and `forceless' equation (1.2).}:
  \begin{equation}
  \frac{d \vec{p}_1}{dt} = q_1 [\vec{E}(RCED)+\vec{\beta}_1 \times (\vec{B}(RCED)]
 \end{equation}
  The fields so-obtained are compared below with the CEM force fields of a charge in uniform motion, derived
  from retarded Li\'{e}nard-Wiechert~\cite{LW} potentials, in the three reference frames introduced in the
  previous section.
    \par \underline{S$^*$, source charge at rest}
  \begin{eqnarray}
  \vec{E}^*(RCED) & = &\frac{Q}{r^2}[\hat{\imath}\cos \psi + \hat{\jmath}\sin \psi] = \frac{Q \vec{r}}{r^3} \\ 
  \vec{E}^*(CEM) & = &\frac{Q}{(r^*)^2}[\hat{\imath}\cos \psi^* + \hat{\jmath}\sin \psi^*] = \frac{Q \vec{r}^*}{(r^*)^3} \\ 
   \vec{B}^*(RCED) & = & \vec{B}^*(CEM) = 0 
   \end{eqnarray}
    \par \underline{S, source and test charge in motion}
  \begin{eqnarray}
    \vec{E}(RCED) & = & \frac{Q}{r^2}\left[\frac{\hat{\imath}\cos \psi}{\gamma_u} + \hat{\jmath} \gamma_u 
  \sin \psi\right] \\
    \vec{B}(RCED) & = & \frac{Q \gamma_u \beta_u \hat{k}\sin \psi}{r^2} \\
  \vec{E}(CEM) & = &\frac{Q[\hat{\imath}\cos \psi + \hat{\jmath}\sin \psi]}{r^2 \gamma_u^2(1-\beta_u^2 \sin^2
   \psi)^\frac{3}{2}} = \frac{Q \vec{r}}{r^3 \gamma_u^2(1-\beta_u^2 \sin^2 \psi)^\frac{3}{2}} \\
   \vec{B}(CEM) & = &  \frac{Q \beta_u \hat{k} \sin \psi}{r^2 \gamma_u^2(1-\beta_u^2 \sin^2 \psi)^\frac{3}{2}}
   \end{eqnarray}
  \par \underline{S', test charge at rest}  
  \begin{eqnarray}
    \vec{E}'(RCED) & = & \frac{Q}{r^2}\left[\frac{\hat{\imath}\cos \psi}{\gamma_w} + \hat{\jmath} \gamma_w
   \sin \psi\right] \\
    \vec{B}'(RCED) & = & -\frac{Q \gamma_w \beta_w \hat{k}\sin \psi}{r^2} \\
  \vec{E}'(CEM) & = &\frac{Q[\hat{\imath}\cos \psi' + \hat{\jmath}\sin \psi']}{(r')^2 \gamma_w^2(1-\beta_w^2 \sin^2
   \psi')^\frac{3}{2}} = \frac{Q \vec{r}'}{(r')^3 \gamma_w^2(1-\beta_w^2 \sin^2 \psi')^\frac{3}{2}} \\
   \vec{B}'(CEM) & = &  -\frac{Q \beta_w \hat{k} \sin \psi'}{(r')^2 \gamma_w^2(1-\beta_w^2 \sin^2 \psi')^\frac{3}{2}}
   \end{eqnarray}   
      The space-time transformation laws of the fields can be read off from Eqns(3.2)-(3.12) with the results:
  \par \underline{RCED fields}
  \begin{eqnarray}
 E_x & = & \frac{E_x^*}{\gamma_u},~~~ E_y  = \gamma_u E_y^*,~~~B_z = \beta_u \gamma_u E_y^* = \beta_u E_y\\
  E'_x & = & \frac{E_x^*}{\gamma_w},~~~ E'_y  = \gamma_w E_y^*,~~~B'_z = -\beta_w \gamma_w E_y^* = - \beta_w E'_y \\
   E'_x & = & \frac{\gamma_u}{\gamma_w}E_x,~~~ E'_y  = \frac{\gamma_w}{\gamma_u} E_y,~~~B'_z = -\frac{\beta_w \gamma_w}
   {\beta_u \gamma_u} B_z
    \end{eqnarray}
   \par \underline{CEM fields}
  \begin{eqnarray}
 E_x & = & \left(\frac{r^*}{r}\right)^2\frac{\cos \psi}{\cos \psi^*}\frac{E_x^*}{\gamma_u^2 f_u},~~~
  E_y = \left(\frac{r^*}{r}\right)^2\frac{\sin \psi}{\sin \psi^*}\frac{E_y^*}{\gamma_u^2 f_u},~~~
  B_z = \beta_u E_y \\
 E'_x & = & \left(\frac{r^*}{r'}\right)^2\frac{\cos \psi'}{\cos \psi^*}\frac{E_x^*}{\gamma_w^2 f_w},~~~
  E'_y = \left(\frac{r^*}{r'}\right)^2\frac{\sin \psi'}{\sin \psi^*}\frac{E_y^*}{\gamma_w^2 f_w},~~~
  B'_z = -\beta_w E'_y \\
 E'_x & = & \left(\frac{r}{r'}\right)^2\frac{\cos \psi'}{\cos \psi}\frac{ \gamma_u^2 f_u E_x}{\gamma_w^2 f_w},~~
  E'_y = \left(\frac{r}{r'}\right)^2\frac{\sin \psi'}{\sin \psi}\frac{ \gamma_u^2 f_u E_y}{\gamma_w^2 f_w},~~
  B'_z = -\frac{\beta_w \gamma_w} {\beta_u \gamma_u} B_z
    \end{eqnarray}
    where
  \[ f_u \equiv (1-\beta_u^2 \sin^2 \psi)^{\frac{3}{2}},~~~f_w \equiv (1-\beta_w^2 \sin^2 \psi')^{\frac{3}{2}}
 \]
   Note that, in the case of the CEM formulae, the possibility of different spatial and angular coordinates in the
   different reference frames is left open even though, as will be shown below,
    the correct application of special relativity
   requires that $r^* = r = r'$ and $\psi^* = \psi = \psi'$, in Fig.1. This point will be returned to later
   when the conventional text-book `relativistic' interpretation of (3.16)-(3.18) is discussed.
    \par Comparison of (3.13)-(3.15) with (3.16)-(3.18) shows that some of the transformation laws of the RCED
    and CEM fields 
    are, apparently, quite different. To investigate this further it is instructive to rederive the 
     transformation laws from
    the defining equations of the electric and magnetic fields in terms of the electromagnetic 4-vector
   potential $A \equiv (A_0;\vec{A})$. These equations are the same in RCED and CEM:
  \begin{eqnarray}
 \vec{E} & \equiv  & -\vec{\nabla}A_0-\frac{1}{c}\frac{\partial \vec{A}}{\partial t} \\
   \vec{B} & \equiv & \vec{\nabla} \times \vec{A}
  \end{eqnarray}
   It is convenient to write these equations in a covariant tensor notation~\cite{JackTN} as:
   \begin{equation}
   F^{\alpha \beta} \equiv \partial^{\alpha} A^{\beta}-  \partial^{\beta} A^{\alpha}
   \end{equation}
   where
   \begin{eqnarray}
    E^i & \equiv & -F^{0i} \\
     B^i & \equiv & -\epsilon_{ijk} F^{jk}  
   \end{eqnarray}
 The greek indices $\alpha, \beta,...$ take the values: $0,1,2,3$ and the latin indices $i,j,...$ the values: $1,2,3$. The alternating tensor $\epsilon_{ijk}$ tates the values +1(-1) for even(odd) permutations of $ijk$. 
   Also 
   \begin{eqnarray}
   \partial^0 & \equiv & \frac{\partial~}{\partial x_0} \equiv \frac{1}{c} \frac{\partial~}{\partial t}  \\
   \partial^i & \equiv & -\frac{\partial~}{\partial x^i}\equiv -\vec{\nabla}
    \end{eqnarray}
    and 
    \[ (x_0=x^0,x^1,x^2,x^3 \equiv ct,x,y,z),~~~~(V^1,V^2,V^3 \equiv  V_x,V_y,V_z) \]
    where $X \equiv (ct;x,y,z)$ is the space-time 4-vector and $\vec{V} \equiv (V_x,V_y,V_z)$ is
    an arbitary 3-vector. $A^{\alpha}$ is, by definition, a 4-vector with the space-time transformations 
    between the frames S and S' give by:
:   \begin{eqnarray}
    A'_x  & = & \gamma_v(A_x-\beta_v A_0) \\
    A'_0  & =  &\gamma_v(A_0-\beta_v A_x)  \\
      A'_y  & = &  A'_y,~~~  A'_z   =   A'_z 
    \end{eqnarray}
    The transformation equations of the electric and magnetic fields will then be determined, via Eqn(3.21), by those 
   of the differential operators $\partial^{\alpha}$. These may be determined from the infinitesimal Lorentz
   transformations:
   \begin{eqnarray}
   dx'  & = & \gamma_v(dx-\beta_v dx_0) \\
    dx'_0  & =  &\gamma_v(dx_0-\beta_v dx)  \\
      dy'  & = &  dy,~~~  dz'  =   dz 
    \end{eqnarray}
    by making some definite hypothesis about the functional space and time dependence of the components of $A^{\alpha}$
     ~\cite{Rosser}. Let $F$ now denote any component of $A^{\alpha}$. 
      If it is assumed that $F$ is analogous to a classical
     field (CF), such as those describing the distribution of velocities in a fluid or the
      temperature at some point $\vec{x}$ in
     a solid body at time $t$, the functional dependence on space and time is simply $F = F(\vec{x},t)$ where
      $x$, $y$ $z$ and $t$ 
     are considered to be independent variables. In this case, the variation of $F$ corresponding to changes in these
     variables is given 
     by the chain rule of differential calculus:
    \begin{eqnarray}
     dF & = & \frac{\partial F}{\partial x} dx+ \frac{\partial F}{\partial y} dy+\frac{\partial F}{\partial z} dz+
          \frac{\partial F}{\partial t} dt \nonumber \\
    & = & \frac{\partial F}{\partial x'} dx'+ \frac{\partial F}{\partial y'} dy'+\frac{\partial F}{\partial z'} dz'+
          \frac{\partial F}{\partial t'} dt'
    \end{eqnarray}
     The relation between the differential operators  $\partial^{\alpha}$ and  $\partial'^{\alpha}$, when operating on $F$, 
    is readily found by using the inverses of (3.29)-(3.31) to eliminate $d\vec{x}$ and $dt$ in the
    second member of  Eqn(3.32) in favour of $d\vec{x}'$ and $dt'$. Collecting terms, equating the coefficients
    of the components of  $d\vec{x}'$ and of $dt'$ in the second and third members of (3.32), and noting that the field
    $F$ is arbitary, gives the following transformation laws for the differential operators:
    \begin{eqnarray}
      \frac{\partial ~}{\partial x'} & = & \gamma_v \left(\frac{\partial ~}{\partial x}
      +\beta_v \frac{\partial ~}{\partial x_0}\right) \\
      \frac{\partial ~}{\partial x'_0} & = & \gamma_v \left(\frac{\partial ~}{\partial x_0}
      +\beta_v \frac{\partial ~}{\partial x}\right) \\
        \frac{\partial ~}{\partial y'} & = & \frac{\partial ~}{\partial y},~~~ 
           \frac{\partial ~}{\partial z'} = \frac{\partial ~}{\partial z} 
     \end{eqnarray}
  These equations imply that  $\partial^{\alpha}$ also transforms as a 4-vector,
  in which case $F^{\alpha \beta}$ is a second rank tensor with the transformation law:
   \begin{equation}
   F'^{\alpha \beta} =  \frac{\partial x'^{\alpha}}{\partial x^{\gamma}} \frac{\partial x'^{\beta}}{\partial x^{\delta}}
       F^{\gamma \delta} 
   \end{equation}
     where repeated upper and lower indices are summed over $0, 1, 2$ and $3$. 
      The transformation equations (3.29)-(3.31) between the frames S and S' give the following values of the 
    transformation coefficients in (3.36):
      \begin{eqnarray}
     \frac{\partial x'}{\partial x} & = & \gamma_v,~~~~\frac{\partial x'}{\partial y} = \frac{\partial x'}{\partial z}=0,~~~~
      \frac{\partial x'}{\partial x_0} = -\gamma_v \beta_v \\
    \frac{\partial x'_0}{\partial x} & = & -\gamma_v \beta_v,~~~~\frac{\partial x_0'}{\partial y}
      =  \frac{\partial x'_0}{\partial z}=0,~~~~\frac{\partial x'_0}{\partial x_0}  =  \gamma_v \\
   \frac{\partial y'}{\partial x} & = & 0,~~~~\frac{\partial y'}{\partial y} = 1,~~~~\frac{\partial y'}{\partial z}  =  0,~~~~
   \frac{\partial y'}{\partial x_0} = 0 \\
   \frac{\partial z'}{\partial x} & = & 0,~~~~\frac{\partial z'}{\partial y} = 0,~~~~\frac{\partial z'}{\partial z}  =  1,~~~~
   \frac{\partial z'}{\partial x_0} = 0
   \end{eqnarray}  
   \par According to (3.36)-(3.40) the transformation law for $E_x$ is:
      \begin{eqnarray}
    E'_x & = & -F'^{01} =-\frac{\partial x'_0}{\partial x^{\gamma}} \frac{\partial x'}{\partial x^{\delta}}
       F^{\gamma \delta}  \nonumber \\
   & = & -\frac{\partial x'_0}{\partial x_0} \frac{\partial x'}{\partial x}
       F^{01}-\frac{\partial x'_0}{\partial x} \frac{\partial x'}{\partial x_0} F^{10} \nonumber \\
   & = & \left(\frac{\partial x'_0}{\partial x_0} \frac{\partial x'}{\partial x}
        -\frac{\partial x'_0}{\partial x} \frac{\partial x'}{\partial x_0}\right)(-F^{01}) \nonumber \\
  & = &  \gamma_v^2(1- \beta_v^2)E_x = E_x
    \end{eqnarray}
  that of $E_y$ is:
      \begin{eqnarray}
    E'_y & = & -F'^{02} =-\frac{\partial x'_0}{\partial x^{\gamma}} \frac{\partial y'}{\partial x^{\delta}}
       F^{\gamma \delta}  \nonumber \\
   & = & -\frac{\partial x'_0}{\partial x_0} \frac{\partial y'}{\partial y}
       F^{02}-\frac{\partial x'_0}{\partial x} \frac{\partial y'}{\partial y} F^{12} \nonumber \\
   & = & -\gamma_v F^{02}+\gamma_v \beta_v F^{12} \nonumber \\
  & = &  \gamma_v(E_y- \beta_v B_z)
    \end{eqnarray}
  and that of $B_z$ is:
      \begin{eqnarray}
    B'_z & = & -F'^{12} =-\frac{\partial x'}{\partial x^{\gamma}} \frac{\partial y'}{\partial x^{\delta}}
       F^{\gamma \delta}  \nonumber \\
   & = & -\frac{\partial x'}{\partial x_0} \frac{\partial y'}{\partial y}
       F^{02}-\frac{\partial x'}{\partial x} \frac{\partial y'}{\partial y} F^{12} \nonumber \\
   & = & \gamma_v \beta_v F^{02}-\gamma_v F^{12} \nonumber \\
  & = &  \gamma_v(B_z- \beta_v E_y)
    \end{eqnarray}
  \par The transformation equations of $E_y$ and $B_z$, (3.42) and(3.43), are consistent with both those
   of the RCED fields given in (3.13)-(3.15) and those of the CEM fields given in (3.16)-(3.18). To see this, consider
   the equation given by combining (3.14) and (3.15) or (3.17) and (3.18):
    \begin{equation}
     E'_y = \frac{\gamma_w }{\gamma_u \beta_u}B_z
     \end{equation}
    The Lorentz transformation of the temporal component of the dimensionless 4-vector velocity ($\gamma_v$;$\gamma_v \beta_v$,0,0)
    gives, in virtue of the velocity addition formula (2.1): 
     \begin{equation}
  \gamma_w = \gamma_u(\gamma_v-\beta_u \gamma_v \beta_v) 
     \end{equation}
   Eqns(3.44) and (3.45) then give:
   \begin{equation}
     E'_y = \gamma_v\left(\frac{B_z}{\beta_u}-\beta_v B_z\right) =  \gamma_v(E_y-\beta_v B_z)
      \end{equation}
  where the last equation in (3.13) or (3.16) has been used, in agreement with (3.42).
  \par Combining the last equation in (3.15) or (3.18) with the  Lorentz transformation of the spatial component of
   ($\gamma_v$;$\gamma_v \beta_v$,0,0):
      \begin{equation}
 \gamma_w \beta_w = \gamma_u(\beta_v \gamma_v-\beta_u \gamma_v)
     \end{equation}  
 gives 
  \begin{equation}
     B'_z = -\gamma_v\left(\frac{\beta_v B_z}{\beta_u}- B_z\right) =  \gamma_v(B_z-\beta_v E_y)
      \end{equation}
   where, again, the last equation in (3.13) or (3.16) has been used, in agreement with (3.43).  
 \par There is, however, a clear contradiction between the transformation law (3.41) of $E_x$ and that of the
   RCED field in (3.15). Whether or not, and on the basis of what assumptions, the CEM transformation equations
    (3.16)-(3.18) are consistent with (3.41)-(3.43) will be discussed at the end of the present section after first 
   explaining the reasons for the difference between (3.41) and the transformation law of the RCED $E_x$ field,
 in the first of equations (3.15).

      \par The essential assumption, leading to the tensor character of $F^{\alpha \beta}$, defined in (3.36)-(3.40),
       is that the components of $A$ are fields, $F$, with space-time functional dependence $F = F(\vec{x},t)$ where
     $\vec{x}$ is the spatial position at which the field is specified
      and $x$, $y$, $z$ and $t$ are independent variables. From this follows the correctness
     of the chain-rule (3.32), and the consequent 4-vector transformation laws (3.33)-(3.35) of $\partial^{\alpha}$.
     Because these laws are the same as those of $A^{\alpha}$, $F^{\alpha \beta}$ has the transformation
     law (3.36), of which Eqn(3.41) is a necessary consequence. The obvious question to ask is, whether the
     components of $A$ are indeed `classical fields' in the sense described above?
     The 4-vector potental, $A$, of a charge in uniform motion in the frame S is given by the following formulae
     in RCED~\cite{JHF1} and CEM~\cite{PPFMC}: 
     \begin{eqnarray}
      A_0(RCED) &  =  & \frac{Q \gamma_u}{r} \\
       \vec{A}(RCED) & = &  \frac{Q \gamma_u \vec{\beta}_u}{r} \\
      A_0(CEM) &  =  &  \frac{Q}{r(1-\beta_u^2\sin^2\psi)^{\frac{1}{2}}} \\
      \vec{A}(CEM) &  =  &  \frac{Q \vec{\beta}_u}{r(1-\beta_u^2\sin^2\psi)^{\frac{1}{2}}} 
     \end{eqnarray}
     where
     \begin{equation}
    r \equiv |\vec{x}_q-\vec{x}_Q|,~~~ \sin \psi = \frac{|y_q-y_Q|}{r}
     \end{equation}
      The RCED and CEM force fields in (3.5)-(3.8) are derived from these potentials, and the definitions (3.19)
      and (3.20), in Appendix A. 
      The retarded Li\'{e}nard-Wiechert potentials~\cite{LW} of CEM are here expressed in their
     `present position' form, so that $r$ in all the equations (3.49) to (3.52) is the instantaneous separation
     of the source and test charges at the instant at which the potentials are evaluated. Two important remarks
     are now in order:
      \begin{itemize}
     \item[(i)] the components of $A$ are {\it not} `classical fields'
                in either RCED or CEM.  
     \item[(ii)] Unlike for RCED, the 4-vector character of the CEM potential is not evident from inspection
                of Eqns(3.51) and (3.52).
      \end{itemize}
      \par In connection with point (i), all the components  of A in both RCED and CEM have a similar
       space-time and source-velocity dependence: 
      \begin{eqnarray}
     F &  = & F( \vec{x}_q,  \vec{x}_Q(t),\beta_u(t))= F( |\vec{x}_q-\vec{x}_Q(t)|,\beta_u(t))
    ~~~({\rm RCED}) \\  
         & = &  F(|\vec{x}_q-\vec{x}_Q(t)|,\sin \psi, \beta_u(t))
    ~~~({\rm CEM})     
      \end{eqnarray}
  which is evidently not that of a classical field:
        \begin{equation}
         F^{CF} = F( \vec{x}_q, t).
      \end{equation}   
    It follows that the chain rule of (3.32), as well as the subsequently derived transformation
     equations (3.33) and (3.34) are of general validity in neither RCED nor CEM. In fact the time dependence
     is, in both cases, implicitly included, for the frames S and S', in that of $\vec{x}_Q$. Since only
    uniform motion is considered in the present section, $\beta_u$ is constant, so all time dependence 
    arises from that of $\vec{x}_Q$. Only in the frame S$^*$, where  $\vec{x}_Q$ is also constant and $\vec{A}$
   vanishes, does the 4-vector potential have the space time functional dependence, (3.56), of a classical field.
    \par The space-time transformation properties of $A$, given the definitions (3.49)-(3.52), are evidently
    determined by those of $\vec{x}_q$ and $\vec{x}_Q$, from which those of the essential parameters $r$ in RCED,
    and $r$ and $\sin \psi$ in CEM, may be determined. As shown in Ref~\cite{JHF1} for RCED, and as will be proved below
     for both RCED and CEM, $r$
     is a Lorentz invariant. This follows in RCED directly from the instantaneous nature of the
    inter-charge force predicted by QED. However, as will now be demonstrated, $r$ must also be a Lorentz scalar
    in the `present position' form (3.51) and (5.52) of the  CEM potentials.  Thus in Fig.1,
    the spatial separation of the source and test charges as well as the angular position of the test charge 
    must be the same in all three reference frames. That this must be
    so, as a consequence of special relativity alone, independently of whether the interaction is 
    instantaneous or retarded, will now be demonstrated.

\begin{figure}[htbp]
\begin{center}
\hspace*{-0.5cm}\mbox{
\epsfysize8.0cm\epsffile{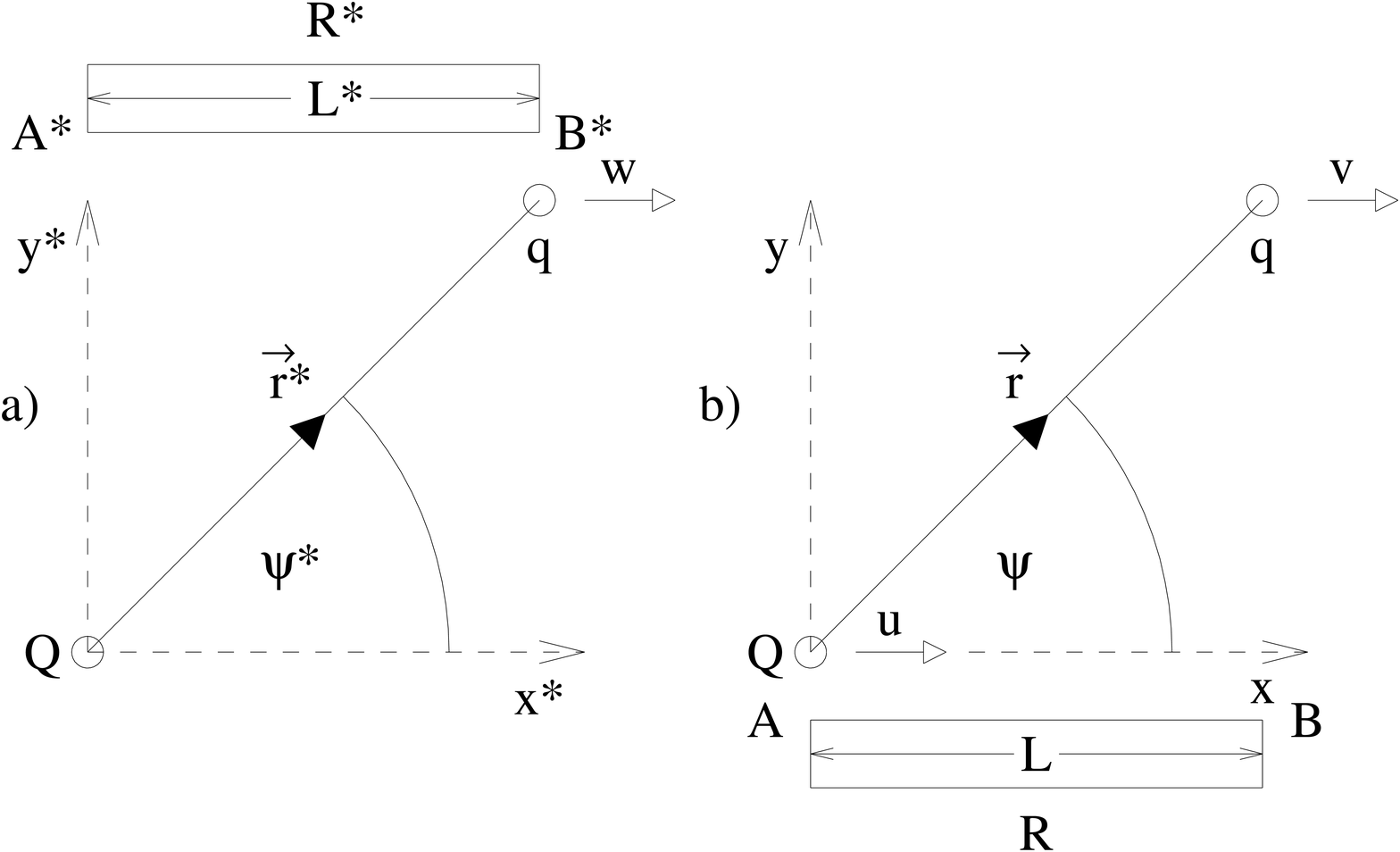}}   
\caption{{\sl a) At the instant $t = t^* = 0$, when the origins of the frames S and S*
  coincide, the ends A* and B* of the ruler R*, of length L*, at rest in S*, have the same
  $x^*$-coordinates as the point-like charges Q and q, respectively. b) At the same instant, the ends A and B
  of the ruler R, of length L, at rest in S, have the same
  $x$-coordinates as Q and q, respectively. Q is at the origin of S*. The LT of transverse 
   coordinate of q gives: $y_q  = y_q^*$, of the longitudinal coordinate: L $=$ L*. See text for discussion.}}
\label{fig-fig2}
\end{center}
\end{figure}

    \par Consider observations in the frames S and S* of the spatial separation of the charges Q and q as shown
     in Fig. 1a and 1b. Suppose that at the instant when $t = t^* = 0$, where $t$, $t^*$ are the times 
     recorded by synchronised clocks at rest in S and S* respectively, the origins of the the frames
     S* (the proper frame of the charge Q) and S coincide. At this instant, as shown in Fig. 2, the rulers
     R and R* of lengths $L$ and $L^*$, at rest in  S and S* respectively, are such that the ends
      A,A*(B,B*) have the same $x,x^*$ coordinates as Q(q). the transformation of the spatial separations
      $r$, $r^*$ of q and Q in S and S* is then given by the transformation of their longitudinal
     separation, equivalent to that of the ruler lengths
       $L$ and $L^*$, since, according to the Lorentz transformation, the transverse separation of
      the charges is invariant:
       \begin{equation}
            y_{\rm q} = r \sin \psi =r^* \sin \psi^* = y^*_{\rm q} 
       \end{equation}
       The following coordinate relations are given by the geometry of Fig. 2:
      \begin{eqnarray}
       x^*_{\rm Q}(0) & = &  x^*_{\rm A^*}(t^*) = 0,~~~~ x^*_{\rm q}(0) =  x^*_{\rm B^*}(t^*) = L^* \\
    x_{\rm Q}(0)& = &  x_{\rm A}(t) = 0,~~~~ x_{\rm q}(0) =  x_{\rm B}(t) = L
      \end{eqnarray}
       The equation of motion of A* in the frame S is
      \begin{equation}
        x_{\rm A^*}(t) = ut
      \end{equation}
     while that of B* is
   \begin{equation}
        x_{\rm B^*}(t) = ut+L
      \end{equation}
   The spatial Lorentz transformations of the world lines of A* and B* are: 
    \begin{eqnarray}
   x^*_{\rm A^*}(t^*)& = & \gamma_u( x_{\rm A^*}(t)-ut) = 0  \\
x^*_{\rm B^*}(t^*)-L^* & = & \gamma_u( x_{\rm B^*}(t)-L-ut) = 0
  \end{eqnarray}
    It follows from (3.61) or (3.63) that 
   \begin{equation}
    L = x_{\rm B^*}(0)
  \end{equation}
  Thus $L$ is a constant, independent of $t$ and $u$, that depends only
   on the choice of coordinate origin in the frame S. In consequence, Eqn(3.63) holds good 
    for all values of $u$, including $u = 0$, in which case, since S* $\rightarrow$ S, 
  $ x^* \rightarrow x$ and $ t^* \rightarrow t$  as  $ u \rightarrow 0$:
   \begin{equation}
    x_{\rm B^*}(t)-L^* =  x_{\rm B^*}(t)-L
 \end{equation}
    so that
   \begin{equation}
    L = L^*
 \end{equation}
    Combining (3.58), (3.59) and (3.65) gives
    \begin{equation}
   L^* = x^*_{\rm B^*}(t^*)-  x^*_{\rm A^*}(t^*) =  x^*_{\rm q}(0)- x^*_{\rm Q}(0) =
     L = x_{\rm B}(t)-x_{\rm A}(t) = x_{\rm q}(0)- x_{\rm Q}(0)
 \end{equation}
   The longitudinal separation of the charges is therefore the same in S and S* at  $t = t^*= 0$,
   ---there is no `length contraction' effect. 
     In virtue of (3.57) and (3.67) it follows that:
        \begin{equation}
   r^* = r = r',~~~ \psi^* =  \psi = \psi' 
   \end{equation}
     as assumed in the space-time transformation formulae of the RCED forces and fields. 
     Since the relations (3.57) and (3.67) hold for all values of $u$, and the transformation
   from S* to S' is obtained from that from S* to S by the replacement $ u \rightarrow -w$. the 
    equations $r^* = r'$ and $\psi^* = \psi'$ in (3.68) also follow directly from (3.57) 
    and (3.67).
    \par An alternative and simpler way to obtain the result $ r^* = r$ is by use of a
     reciprocity relation that exploits the symmetry between the configurations
    shown in Fig.2a and Fig.2b . As stated by Pauli~\cite{Pauli} and applied to the present
    problem:
    \par{\tt The contraction of lengths at rest in S$^*$ and observed in S is equal to \newline that
      of lengths in S and observed in  S$^*$.}
    \par where the `lengths' considered are those parallel to the $x,x^*$ axes.
     \par In the present case the `length at rest in S$^*$' is the length, ${\rm L}^*$ of the ruler R* in the frame in which the source charge is at rest (Fig.2a).
       The length `observed in S' is is the length, ${\rm L}$ of the ruler R in this frame (Fig.2b).
    Similarly the `length at rest in S' 
     is ${\rm L}$, and this length as `observed in S$^*$' is ${\rm L}^*$. The reciprocity relation
     then gives:
     \begin{equation}
     {\rm L}  = \alpha(u){\rm L}^* 
     \end{equation}
     and 
  \begin{equation}
     {\rm L}^* = \alpha(u) {\rm L}
     \end{equation}
   where the contraction factor $\alpha(u)$ is an even function of the velocity $u$ of frame S*  
   relative to the frame S. It follows from (3.69) and (3.70) that:
       \begin{equation}
      {\rm L} = \alpha(u){\rm L}^* =  \alpha(u)^2  {\rm L}
     \end{equation}  
      so that  $\alpha(u) = \pm 1$. Since  ${\rm L}$ and ${\rm L}^* $ are positive quantites it follows
       from (3.69) or (3.70) that  ${\rm L}^* = {\rm L}$  in agreement with Eqn(3.66)

   \par  A consequence of Eqn(3.66) is
    that the x-derivatives in different frames do not transform according to Eqn(3.33) but are
    instead invariant\footnote{Since in Figs. 1 and 2 $x_{\rm Q} =  x^*_{\rm Q} = 0$ 
      and $y_{\rm q} = y^*_{\rm q}$, then ${\rm L} = {\rm L}^*$ implies that $x_{\rm q} = x^*_{\rm q}$
       so that $dx_{\rm q} = dx^*_{\rm q}$.}:
     \begin{equation}
    \frac{\partial~}{\partial x_q^*} = \frac{\partial~}{\partial x_q} =  \frac{\partial~}{\partial x'_q}
   \end{equation}   
    Because of the space-time functional dependence of each component of $A$ in Eqns(3.54) and (3.55),
    the chain rule,
    appropriate for the functional dependence of the classical field in Eqn(3.56),
     cannot be used to derive the transformation
    (3.34) of the time derivative. Instead, it follows from (3.54) or (3.55) that:
      \begin{equation}
    \left. \frac{\partial F}{\partial t}\right|_{x_q} =  \left. \frac{\partial F}{\partial x_Q}\right|_{x_q}
     \frac{d x_Q}{d t} = \left. -u \frac{\partial F}{\partial x_q}\right|_t  
   \end{equation} 
     where $F$ is some arbitary component of a 4-vector potential or force field.
     In consequence,
     \begin{equation}
  \beta_u \frac{\partial A_x}{\partial x_q } = -\frac{1}{c}  \frac{\partial A_x}{\partial t}
    \end{equation} 
    The transformation equations of the 4-vector potential between the frames S and S$^*$ are:
      \begin{eqnarray} 
     A_0^* & = & \gamma_u (A_0-\beta_u A_x) \\
     A_x^* & = &\gamma_u (A_x-\beta_u A_0) = 0\\
     A_y^* & = &  A_y = 0 \\
  A_z^* & = &  A_z = 0
  \end{eqnarray}
   Using the definition (3.19) of the electric field (3.72), (3.73) and (3.75) give:
      \begin{eqnarray}
     E_x^* & = &  - \frac{\partial A_0^*}{\partial x_q^*} = -\gamma_u \left( \frac{\partial A_0}{\partial x_q}
        -\beta_u \frac{\partial A_x}{\partial x_q}\right)\nonumber \\
  & = &  -\gamma_u \left( \frac{\partial A_0}{\partial x_q}
        +\frac{1}{c}\frac{\partial A_x}{\partial t}\right) \nonumber \\
   & = & \gamma_u E_x 
   \end{eqnarray}
  This equation, transposed, is the first one of (3.13). 
   The y-component of the electric field in S$^*$ is given by (3.19) as:
  \begin{equation}
   E_y^* = - \frac{\partial A_0^*}{\partial y_q^*}
  \end{equation}
  The Lorentz invariance of $y_q$, Eqn(3.57), and Eqn(3.75) give:
  \begin{eqnarray} 
   E_y^* & = & - \gamma_u\left( \frac{\partial A_0}{\partial y_q}
        -\beta_u \frac{\partial A_x}{\partial y_q}\right) \nonumber  \\
     & = & - \gamma_u(1-\beta_u^2)\frac{\partial A_0}{\partial y_q} \nonumber \\
       & = & \frac{E_y}{\gamma_u} 
   \end{eqnarray}
   in agreement with the second of Eqns(3.13).
   Finally, using (3.23) and (3.76):
    \begin{equation}
   B_z = - \frac{\partial A_x}{\partial y_q} = -\beta_u  \frac{\partial A_0}{\partial y_q} = \beta_u E_y
   \end{equation}
   which is the last of Eqns(3.13)
    The field transformations between S$^*$ and S', Eqns(3.14) are given by the replacement $u \rightarrow -w$ in 
     (3.79) and(3.81). Those between S and S', Eqns(3.15), are derived directly from (3.13) and (3.14).

     \par It remains to discuss the consistency, or otherwise, of the transformation laws of the CEM fields,
      in (3.16)-(3.18), with those of the RCED fields in (3.13)-(3.15) just considered, and those of the second
      rank tensor $F^{\alpha \beta}$ in (3.41)-(3.43). The standard text book treatment, which claims consistency
      between (3.16)-(3.18) and (3.41)-(3.43) will first be recalled, and then
      shown to be based on an incorrect application of special relativity.
      In the following, for definitness, the presentation in Jackson's book~\cite{JackFMC} is closely followed.
      Similar reasoning may be found in many other text books on classical electromagnetism~\cite{Purcell,PPFMC,LLFMC}.
    In the notation of the present paper, the field point (or the position of the test
      charge $q$) is at rest at the origin of S' while the source charge is at rest at the origin of the S$^*$
      frame. The $ x^*$ and $x'$ axes are parallel but separated by the 
      distance $\Delta y^* = \Delta y'$. The space and time coordinates in both S' and S$^*$ are considered.
      At $t' = t^* = 0$ the test and source charge have the same x-coordinate in  both frames: $x' = x^* = 0$.
      At the later times $t'$,  $t^*$ they therefore have an x-separations of $\Delta x' \equiv x'_q-x'_Q = w t'$ in S',
      and  $\Delta x^* \equiv x^*_q-x^*_Q = w t^*$ in S$^*$. 
       Three relativistic ans\"{a}tze are then introduced~\cite{JackFMC}. The
  first  step is to apply the Lorentz
       transformation of time between S' and S$^*$ to the space-time coordinates of the test charge:
     \begin{equation}
       t^* = \gamma_w(t'+\frac{\beta_w x'_q}{c})
     \end{equation}
    Since $x'_q = 0$ at all times this gives the first ansatz A1
     \footnote{In Ref.~\cite{JackFMC} (see Fig.11.8) the test charge is at rest in the unprimed frame, whereas the
       primed frame is the rest frame of the source charge, so that A1 is written: $t' = \gamma t$.}:
     \begin{equation}
        t^* = \gamma_w t'~~~~{\rm A1}
     \end{equation}
    The second and third ans\"{a}tze, A2, and A3 are the relations giving the separation of the source and test charge
    in S' and  S$^*$ repectively:
          \begin{equation}
           \Delta x' = w t'~~~~{\rm A2}
     \end{equation}
          \begin{equation}
           \Delta x^* = w t^*~~~~{\rm A3}
     \end{equation}
     where it is assumed that $ t'$ and $t^*$ are related by (3.84). 
  Combining (3.83), (3.84) and (3.85) 
     leads to the relation:
     \begin{equation}
       \Delta x^* =  w t^* = \gamma_w w t' = \gamma_w  \Delta x'
     \end{equation}
    Thus the magnitude of the x-interval between the charges is found to be contracted by the factor $1/\gamma_w$ in the S'
    frame.
    This last equation is now used to relate $r'$ to $r^*$:
  \begin{eqnarray} 
   (r^*)^2 & = & (\Delta x^*)^2+(\Delta y^*)^2 =  \gamma_w^2(\Delta x')^2+(\Delta y')^2 \nonumber \\
    & = & (r')^2\left[1+ (\gamma_w^2-1)\left(\frac{\Delta x'}{r'}\right)^2\right] =
        (r')^2(1+\gamma_w^2 \beta_w^2 \cos^2 \psi') \nonumber \\ 
    & = &(r')^2 \gamma_w^2(1-\beta_w^2 \sin^2  \psi')
   \end{eqnarray}
   as well as $\psi^*$ and $\psi'$:
  \begin{eqnarray}
   \sin \psi' & = & \frac{\Delta y'}{r'} = \left(\frac{r^*}{r'}\right) \frac{\Delta y^*}{r^*}
    =  \left(\frac{r^*}{r'}\right)\sin \psi^* \\
  \cos \psi' & = & -\frac{\Delta x'}{ r'} = \left(\frac{r^*}{ \gamma_w r'}\right) \frac{\Delta x^*}{r^*}
    =  \left(\frac{r^*}{ \gamma_w r'}\right)\cos \psi^*
   \end{eqnarray}
   Substitution of (3.87) and(3.89) into the first equation in (3.17) gives:
     \begin{equation}
        E'_x = E_x^*
      \end{equation}
  in agreement with (3.41), but not with the RCED transformation in Eqns(3.14).
   Eqns(3.87) and (3.88) together with the second equation in (3.17) give:
     \begin{equation}
        E'_y = \gamma_w E_y^*
      \end{equation}
   in agreement both with the RCED transformation in Eqns(3.14) and the tensor transformation
   (3.42)\footnote{ The magnetic field transformation between the frames S$^*$ and S' is given by the replacements
          $v \rightarrow w$, $ E_y \rightarrow E_y^*$ and $ B_z \rightarrow B_z^* = 0$ in Eqn(3.42).}.
 \par The three ans\"{a}tze  A1 (Eqn(3.83)) and A2 (Eqn(3.84)) and A3 (Eqn(3.85)), on which the conventional analysis
   of the transformation properties of the CEM fields is based, are now critically examined.
   Their combination leads to the relation (3.87) which is in contradiction with
   the Lorentz scalar nature, $r^* = r'$, of the intercharge separation derived above. The crucial point 
     is that no distinction  is made between
     the `frame time' (i.e. that registered by all synchronised clocks at rest in the frame, at any instant,
        and viewed by an observer at rest in the frame),
     with the apparent times of clocks when viewed by an observer in motion relative to the proper frames
     of the clocks. Only the frame times $t'$ and $t^*$ are relevant for the relations $ \Delta x' = w t'$
     and $\Delta x^* = w t^*$. In fact, the transformation equation (3.83) (and its reciprocal) relate not two frame
     times , but rather, in each case, a frame time to an apparent time. To understand the actual physical meaning
    of the time transformation equation (3.83) it is instructive to
   apply the same transformation to the source rather than the test charge.
    Making use of the relation $x'_Q = -w t'$ (see Fig.1c)) it is found that:
     \begin{equation}
     t' = \gamma_w (t^*)^{app} 
      \end{equation}
     where $ (t^*)^{app}$ is the apparent time of a clock moving together with the source charge as viewed by an
    observer at rest in S' and $t'$ is the S' frame time.
    In a similar way, the transformation far an experiment reciprocal
    to the one described by Eqn(3.83):
       \begin{equation}
       t' = \gamma_w(t^*-\frac{\beta_w x_q^*}{c})
     \end{equation} 
    can be used to calculate the apparent time of  a clock moving together with the test charge as viewed by an
    observer at rest in S$^*$. In this case $x_q^* = w t^*$ (see Fig.1a) and (3.94) gives:
      \begin{equation}
     t^* = \gamma_w (t')^{app} 
      \end{equation}
     Here $ t^*$ is the S$^*$ frame time.
    Both (3.93) and (3.95) are examples of the well-known time dilatation effect.
   In (3.93) a moving clock with S$^*$
    as proper frame is viewed from S'. In (3.95) a moving clock with S'
    as proper frame is viewed from S$^*$. Comparing now (3.95) with (3.84) it is clear that if $t^*$ in (3.84)
     and (3.95) is the frame time in S$^*$ then $t'$ in (3.84) is {\it not} the frame time in S' but rather the
     apparent time
     $(t')^{app}$ of a clock at rest in S' as viewed from S$^*$. Thus the ansatz A1 is false, if the value
      of $t'$ in it is identified with the frame time in the correct ansatz A2. Thus although A2 and A3
    correctly define the relations between $\Delta x'$, $t'$ in a
    primary experiment and $\Delta x^*$, $t^*$ in its reciprocal, respectively, 
     A1 relates an apparent time to a frame time, not the two frame times. This means that, although the 
    first member of Eqn(3.87) is correct, the 
    last one, derived by combining all three ans\"{a}tze, is not. 
     \par Indeed, if instead of making the substitution  $(t')^{app} \rightarrow t'$ in (3.95), leading to (3.84).
 the substitution  $(t^*)^{app} \rightarrow t^*$ is made in (3.92) the relation $t' = \gamma_w t^*$ is obtained
 instead of (3.84). Combining with (3.85) then leads to the relation $\Delta x' = \gamma_w \Delta x^*$ instead
   of (3.87), so that the $x$-interval is {\it elongated} rather than contracted in the frame S'.
   Because of the evident spatial symmetry of the source and test charges this derivation is on
   exactly the same logical footing (and equally false!) as the derivation of (3.87).  

     Since, in fact, no observations of moving clocks are performed in the problem presently under discussion
     the time transformation equation (3.83) has, in any case, no possible relevance to the discussion.
     What is important,
     as analysed above, leading to Eqn(3.67),
     are the spatial separations of the charges at corresponding instants in the two reference frames.

       \par Substituting now the correct spatial separations and angles from (3.67) into
      (3.17) leads to the transformation laws of the CEM fields;
       \begin{equation}
          E'_x(CEM) = \frac{E_x^*(CEM)}{\gamma_w^2 f_w}
        \end{equation}
      and 
     \begin{equation}
          E'_y(CEM) = \frac{E_y^*(CEM)}{\gamma_w^2 f_w}
        \end{equation}
       in contradiction to both the transformation laws of the RCED field and those, Eqns(3.41) and (3.42), of 
       a classical field transforming as a second rank tensor. 
        \par Finally in this section the transformation laws the electromagnetic potential in RCED and CEM 
          are discussed. Since $r = r^*$ Eqns(3.49)-(3.52) may be written as:
             \begin{eqnarray}
      A_0(RCED) &  =  & \frac{Q \gamma_u}{r}  = \gamma_u A_0^* \\
       \vec{A}(RCED) & = &  \frac{Q \gamma_u \vec{\beta}_u}{r} =   \gamma_u  \vec{\beta}_u  A_0^* \\
      A_0(CEM) &  =  &  \frac{Q}{r(1-\beta_u^2\sin^2\psi)^{\frac{1}{2}}} =  \frac{A_0^*}{(1-\beta_u^2\sin^2\psi)^{\frac{1}{2}}} \\
  \vec{A}(CEM) &  =  &  \frac{\vec{\beta}_u A_0^*}{(1-\beta_u^2\sin^2\psi)^{\frac{1}{2}}} 
     \end{eqnarray}
      so that $A(RCED)$, but not $A(CEM)$, transforms as a 4-vector. On the basis of the ans\"{a}tze A1, A2 and A3 the
   the equation relating $r$ and $r^*$ analogous to
   (3.88)\footnote{This equation is given by the replacements: $w
   \rightarrow u$, $\psi' \rightarrow \psi$  and $r' \rightarrow r$ in (3.88).} may be
    used to write:
             \begin{eqnarray}
      A_0(CEM) &  =  &  \frac{Q}{r(1-\beta_u^2\sin^2\psi)^{\frac{1}{2}}} =  \frac{Q}{r^*}\left(\frac{r^*}{r}\right)\frac{1}
      {(1-\beta_u^2\sin^2\psi)^{\frac{1}{2}}} = \gamma_u  A_0^* \\
  \vec{A}(CEM) &  =  &   \frac{Q}{r^*}\left(\frac{r^*}{r}\right)\frac{\vec{\beta}_u}
     {(1-\beta_u^2\sin^2\psi)^{\frac{1}{2}}} = \gamma_u \vec{\beta}_u A_0^*  
     \end{eqnarray}
      In this case $A(CEM)$ transforms as a 4-vector. However, as explained above, the anstatz A1 from which (3.87) is derived 
     is
     false. In conclusion, the `present position'
    retarded Li\'{e}nard-Wiechert potentials (3.51) and (3.52), when
     the correct relativistic transformation (3.68) is applied to the intercharge
     separation, do not, unlike the RCED ones (3.49) and (3.50), constitute
     a 4-vector.
    
 \SECTION{\bf{Breakdown of the electrodynamic Maxwell equations in free space for the force fields of RCED}}
    Maxwell's equations played a very important role in the development of the theory of special relativity. The possibility
    of the non-absolute character of time was suggested by the space and time transformations introduced by Voigt~\cite{Voigt}
    \footnote{The Voigt transformation differs from the Lorentz transformation of Eqns(3.62)-(3.65) by an overall
    multiplicative factor $1/\tilde{\gamma}$ on the right sides of the equations.} in order to leave
     the wave equation invariant in different
     inertial frames. The space-time Lorentz transformation
      was originally obtained
      by Larmor \footnote{The Lorentz Transformation (1)-(4)
   was first given by Larmor in 1900~\cite{Larmor}. In this paper the proposed electric
     and magnetic field transformations are the same as in Eqns(3.41)-(3.43).  The same transformations
  appear, without citing Larmor's prior publication, in 1904, in Lorentz's last pre-relativity paper~\cite{Lorentz}. The latter paper was cited
  in 1905 by Poincar\'{e}~\cite{Poincare1905} who first introduced the appellation `Lorentz Transformation'.
  See Ref.~\cite{Kittel} for a discussion of historical priority issues} from the requirement that Maxwell's
    equations should remain invariant under this transformation. As will be recalled below, this also requires that electric
    and magnetic fields transform according to Eqn(3.36), that is, as classical fields that are components of a second rank tensor.
     The equation for `electromagnetic waves' may be derived by combining three of the free space Maxwell equations. The invariance
    of this wave equation for different inertial frames shows that the speed, $c$, of the `waves' is the same in all inertial
     frames. This was the great triumph of Maxwell's electromagnetic field theory. It was clearly the primary motivation
    for Einstein to promote the constancy of the speed of light (which was identified with that of the `electromagnetic waves')
     to the status of a 
     fundamental postulate on which the theory of relativity was based.
  Indeed, since the relativity principle itself is equally valid in Newtonian physics, it is only this
   second postulate of Einstein which differentiates special from Galilean relativity.
     \par The above account describes the historical development of the subject as recounted in text books on classical 
      electrodynamics and special relativity. However, there is a very serious inconsistency in this line of argument.
      Although it has, until now, been universally assumed that the force fields of classical electrodynamics discussed
    in the previous section do satisfy Maxwell's equations, they are manifestly {\it not} classical fields with
    space-time functional dependence $F^{CF} = F(\vec{x}_q,t)$ where $x_q$, $y_q$, $z_q$ and $t$ are independent variables.
    In fact, the components of electric and magnetic fields have a space-time functional dependence
      similar to a component of the corresponding 4-vector potentials as given in (3.54) for RCED and (3.55) for CEM.
      As will be shown below, a consequence of this is that, although the Faraday-Lenz law and the magnetic field divergence
     equations are satisfied by all the force fields considered in the previous section, as consequence of 3-vector
      identities, this is not true for all components of the electrodynamic Amp\`{e}re's law, and the electric field divergence
     equation, for the case of the RCED fields. If these latter equations are not obeyed by the fields, Maxwell's derivation of
    `electromagnetic waves'
     breaks down for them. The interesting question then is whether some other type of field (classical or otherwise)
     {\it does}
     satisfy a wave equation and so may be associated with the propagation of light. This question will be addressed 
     in the following section. However, it may be already stated here that the answer is in the affirmative.
     \par In recalling Maxwell's derivation of  `electromagnetic waves' it will be found convenient to introduce
     a special notation, calligraphic symbols $\vec{\E}$ and $\vec{\B}$ to denote hypothetical classical fields with 
     space-time functional dependence $\vec{\E} = \vec{\E}(\vec{x},t)$,  $\vec{\B} = \vec{\B}(\vec{x},t)$. Such fields
     transform according
     to (3.36) with corresponding space and time derivatives transforming according to (3.33) to (3.35). It will also be assumed,
    in the following, that these fields satisfy the free-space Maxwell equations:
     \begin{eqnarray}
      \vec{\nabla}\cdot \vec{\E} & = & 0 \\
     \vec{\nabla}\cdot \vec{\B} & = & 0 \\
    \vec{\nabla} \times \vec{\E} +\frac{1}{c}\frac{\partial\vec{\B}}{\partial t} & = & 0 \\
 \vec{\nabla} \times \vec{\B} -\frac{1}{c}\frac{\partial \vec{\E}}{\partial t} & = & 0
     \end{eqnarray}
    The Lorentz covariance of (4.3) (The Faraday-Lenz law) and (4.4) (Amp\`{e}re's law) are readily demonstrated using
     (3.33)-(3.36). Two examples containing fields similar to those generated by a uniformly moving charge where $\B_x =0$
     will be worked out. The first is the z-component of (4.3):
   \begin{equation}
     \frac{\partial \E_y}{\partial x}-  \frac{\partial \E_x}{\partial y}+ \frac{1}{c} \frac{\partial \B_z}{\partial t} = 0
    \end{equation}
  Use of the inverses of (3.33)-(3.35) and (3.41)-(3.43) enables (4.5) to be written as:
     \begin{eqnarray}
   &   & \gamma_v\left(\frac{\partial~}{\partial x'}-\frac{\beta_v}{c}\frac{\partial~}{\partial t'}\right)
            \gamma_v(\E'_y+\beta_v \B'_z) - \frac{\partial \E'_x}{\partial y'}  \nonumber \\
   &   &   + \gamma_v\left(\frac{1}{c} \frac{\partial~}{\partial t'}- \beta_v\frac{\partial~}{\partial x'}\right)
            \gamma_v(\B'_z+\beta_v \E'_y)  \nonumber \\
   & = & \gamma_v^2\left(\frac{\partial \E'_y}{\partial x'} +\beta_v\frac{\partial \B'_z}{\partial x'}
    -\frac{\beta_v}{c}\frac{\partial \E'_y}{\partial t'}  -\frac{\beta_v^2}{c}\frac{\partial \B'_z}{\partial t'} \right)
   - \frac{\partial \E'_x}{\partial y'} \nonumber \\ 
   &  & +\gamma_v^2\left(\frac{1}{c} \frac{\partial \B'_z}{\partial t'} -\beta_v\frac{\partial \B'_z}{\partial x'}    
    +\frac{\beta_v}{c}\frac{\partial \E'_y}{\partial t'}  - \beta_v^2 \frac{\partial \E'_y}{\partial x'} \right) = 0
     \nonumber
   \end{eqnarray}
    or, collecting terms
     \begin{eqnarray}
   &   & \gamma_v^2(1-\beta_v^2)\frac{\partial \E'_y}{\partial x'}-\frac{\partial \E'_x}{\partial y'}
       + \gamma_v^2(1-\beta_v^2)\frac{1}{c}\frac{\partial \B'_z}{\partial t'} \nonumber \\
   & = &\frac{\partial \E'_y}{\partial x'}-\frac{\partial \E'_x}{\partial y'}
      +\frac{1}{c}\frac{\partial \B'_z}{\partial t'} = 0
    \end{eqnarray}
  demonstrating the covariance of (4.5).
  The second example is the x-component of (4.4):
  \begin{equation}
     \frac{\partial \B_z}{\partial y}-  \frac{\partial \B_y}{\partial z} -\frac{1}{c} \frac{\partial \E_x}{\partial t} = 0
    \end{equation}    
     Using the inverses of (3.33)-(3.35), (3.41) and (3.43) as well as the relation similar to the inverse of (3.43), derived
     in a similar manner: $\B_y = \gamma_v(\B'_y-\beta_v \E'_z)$, (4.7) may be written as:
     \begin{eqnarray}
 &  & \frac{\partial ~}{\partial y'} \gamma_v(\B'_z+\beta_v\E'_y) - 
       \frac{\partial ~}{\partial z'} \gamma_v(\B'_y-\beta_v\E'_z)-
        \gamma_v \left(\frac{1}{c} \frac{\partial \E'_x}{\partial t'} - \beta_v \frac{\partial \E'_x}{\partial x'}\right)
   \nonumber \\
  &  =  & \gamma_v\left[  \frac{\partial \B'_z}{\partial y'}- \frac{\partial \B'_y}{\partial z'} -
         \frac{1}{c} \frac{\partial \E'_x}{\partial t'}+ \beta_v \vec{\nabla}'\cdot \vec{\E}'\right]   \nonumber \\
    &  =  &  \gamma_v\left[  \frac{\partial \B'_z}{\partial y'}- \frac{\partial \B'_y}{\partial z'} -
         \frac{1}{c} \frac{\partial \E'_x}{\partial t'}'\right] = 0
    \end{eqnarray}
  where, in the last line, the Maxwell equation (4.1) is assumed to hold in the S' frame. As will be shown below, 
  (4.1), is valid for the RCED field only in the frame S$^*$ where the source
   charge is at rest, i.e. for an electrostatic field. The calculations just performed show that the covariance of (4.7) 
   requires also that of the Maxwell equation (4.1), whereas no such requirement is needed to demonstrate the covariance
   of (4.5). Indeed, since the Faraday-Lenz law is an identity following directly from the definitions, (3.19) and (3.20)
   of the electric and magnetic fields, it must be valid in any frame and hence be covariant. To show this it is only
   necessary to take the curl of (3.19):
     \begin{equation}
  \vec{\nabla} \times \vec{E} = - \vec{\nabla} \times (\vec{\nabla} A_0) - \frac{1}{c}
      \frac{\partial ( \vec{\nabla} \times \vec{A})}{\partial t} = - \frac{1}{c} \frac{\partial \vec{B}}{\partial t}
    \end{equation}
   since $ \vec{\nabla} \times (\vec{\nabla}\phi) \equiv 0$, for an arbitary scalar function $\phi$,  and where the definition (3.20) of the magnetic field 
   has been used. Eqn(4.9) is then valid whatever is the space-time functional dependence of the fields $\vec{E}$
    and $\vec{B}$. In contrast there is no such 3-vector identity underlying Amp\`{e}re's law (4.4). 
    \par The wave equation for the classical field $\vec{\E}$ is derived by taking the curl of (4.3) and the time derivative 
     of (4.4):
     \begin{eqnarray}
    &  &  \vec{\nabla} \times (\vec{\nabla} \times \vec{\E})+\frac{1}{c}
      \frac{\partial ( \vec{\nabla} \times \vec{\B})}{\partial t}    \nonumber \\
    & = &  \vec{\nabla} (\vec{\nabla} \cdot \vec{\E}) -  \vec{\nabla}^2 \vec{\E}  +\frac{1}{c}
      \frac{\partial ( \vec{\nabla} \times \vec{\B})}{\partial t}    \nonumber \\
    & = &   -  \vec{\nabla}^2 \vec{\E} +\frac{1}{c}
      \frac{\partial ( \vec{\nabla} \times \vec{\B})}{\partial t} = 0 
    \end{eqnarray}
      \begin{equation}
       \frac{\partial ( \vec{\nabla} \times \vec{\B})}{\partial t} -\frac{1}{c} \frac{\partial^2 \vec{\E}}{\partial t^2}
        = 0
      \end{equation}
      where, in the last line of (4.10) the electric field divergence equation (4.1) has been used.
     Combining (4.10) and (4.11) gives the wave equation
       \begin{equation}
       \vec{\nabla}^2 \vec{\E} -\frac{1}{c^2} \frac{\partial^2 \vec{\E}}{\partial t^2}
        = 0
      \end{equation}
     A similar wave equation for $\vec{\B}$ is obtained by combining the curl of (4.4) with the time derivative of (4.3).
    
     \par Since the divergence equation of the magnetic field, (4.2) may be written, using the magnetic field definition
   (3.20), as a triple scalar product of 3-vectors:
       \begin{equation} 
     \vec{\nabla} \cdot (\vec{\nabla} \times \vec{\A}) = 0 
          \end{equation}     
      it vanishes for any space-time functional dependence of $ \vec{\A}$ in virtue of the identity, valid for any
      3-vectors $\vec{a}$ and $\vec{b}$: $\vec{a} \cdot (\vec{a} \times \vec{b}) \equiv 0$. Again, no such identity exists
       to ensure the validity of the electric field divergence equation (4.1). 
         \par The validity, or not, of the Maxwell equation (4.1) that is not guaranteed by a 3-vector identity, is now
        tested by direct substitution, in this equation, of the fields in (3.5)-(3.8). This straightforward, but somewhat 
      lengthy, calculation is presented in Appendix B. The following results are obtained.
       \begin{eqnarray}
        \vec{\nabla} \cdot \vec{E}(RCED)& = &\frac{Q}{r^3}\left(\gamma_u-\frac{1}{\gamma_u}\right)(2-3\sin^2\psi) \\
     \vec{\nabla} \cdot \vec{E}(CEM) & = & 0
        \end{eqnarray}
       The corresponding equations in the frame S' are:
      \begin{eqnarray}
        \vec{\nabla}' \cdot \vec{E}'(RCED)& = &\frac{Q}{r^3}\left(\gamma_w-\frac{1}{\gamma_w}\right)(2-3\sin^2\psi) \\
     \vec{\nabla}' \cdot \vec{E}'(CEM) & = & 0
        \end{eqnarray}
      The electric field divergence equation is therefore covariant for the CEM field but not for the RCED one.
     \par In order to discuss the validity of Amp\`{e}re's law, (4.4) for the RCED and CEM fields it is convenient
      to perform a rotation by an angle $\phi$ around the x-axes of the coordinate systems used in Eqns(3.2)-(3.12).
        This gives, for example, in the S system:
   \begin{eqnarray}
    \vec{E}(RCED) & = & \frac{Q}{r^2}\left[\frac{\hat{\imath}\cos \psi}{\gamma_u} + \gamma_u 
  \sin \psi(\hat{\jmath} \cos \phi +\hat{k} \sin \phi)\right] \\
    \vec{B}(RCED) & = & \frac{Q \gamma_u \beta_u \sin \psi(\hat{k} \cos \phi-\hat{\jmath} \sin \phi)}{r^2} 
   \end{eqnarray}    
  Inspection of these equations shows that:
   \begin{equation}
     B_z = \beta_u E_y,~~~ B_y = -\beta_u E_z
   \end{equation} 
   The same relations hold for the CEM fields.
   Use of these equations and the equation, (3.73) relating the $x$- and $t$-derivatives, also valid for both the 
   RCED and CEM fields, enables the x-component of Amp\`{e}re's law to be derived directly from the electric field
   divergence equation (4.1) with the replacement $\vec{\E} \rightarrow \vec{E}$. For the CEM fields:
     \begin{eqnarray}
     \vec{\nabla} \cdot \vec{E}(CEM) & = & \frac{\partial E_x}{\partial x_q}+ \frac{\partial E_y}{\partial y_q}
         +\frac{\partial E_z}{\partial z_q} \nonumber \\
  & = &   \frac{1}{\beta_u}\left[ \beta_u \frac{\partial E_x}{\partial x_q}+ \frac{\partial(\beta_u E_y)}{\partial y_q}
          +\frac{\partial(\beta_u E_z)}{\partial z_q}\right] \nonumber \\  
 & = &   \frac{1}{\beta_u}\left[ -\frac{1}{c} \frac{\partial E_x}{\partial t}+ \frac{\partial  B_z}{\partial y_q}
          -\frac{\partial  B_y}{\partial z_q}\right] = 0
\end{eqnarray}
   so that
  \par \underline{ CEM fields} 
 \begin{equation}
  \frac{\partial  B_z}{\partial y_q} -\frac{\partial  B_y}{\partial z_q}
   -\frac{1}{c} \frac{\partial E_x}{\partial t} = 0
 \end{equation}
 which is the x-component of (4.4) with the replacements  $\vec{\E} \rightarrow \vec{E}$ and  $\vec{\B} \rightarrow \vec{B}$.
  \par Repeating the calculation shown in (4.21) for RCED fields and using (4.14) shows that the RCED fields do not satisfy Amp\`{e}re's law 
   but rather give the relation:
  \par \underline{ RCED fields} 
 \begin{equation}
  \frac{\partial  B_z}{\partial y_q} -\frac{\partial  B_y}{\partial z_q}
   -\frac{1}{c} \frac{\partial E_x}{\partial t} = \frac{Q \beta_u}{r^3}\left(\gamma_u-\frac{1}{\gamma_u}\right)(2-3\sin^2\psi)
 \end{equation}
  The x-component of Amp\`{e}re's law is therefore not covariant for the RCED fields.
  \par Since $B_x = 0$ in both RCED and CEM, for the case of a source charge in uniform motion
    parallel to the x-axis, the y- and z-components of Amp\`{e}re's law simplify to:
 \begin{eqnarray}
  -\frac{\partial  B_z}{\partial x_q}-\frac{1}{c}\frac{\partial  E_y}{\partial t} & = & 0   \\ 
 \frac{\partial  B_y}{\partial x_q}-\frac{1}{c}\frac{\partial  E_z}{\partial t} & = & 0
 \end{eqnarray}
 Using (4.20), (4.24) may be written as:
 \begin{equation}
 \frac{\partial  B_z}{\partial x_q} = \beta_u \frac{\partial  E_y}{\partial x_q} = -\frac{1}{c}\frac{\partial  E_y}{\partial t}
 \end{equation}
  Similarly, (4.25) may be written as:
 \begin{equation}
 -\frac{\partial  B_y}{\partial x_q} = \beta_u \frac{\partial  E_z}{\partial x_q} = -\frac{1}{c}\frac{\partial  E_z}{\partial t}
 \end{equation}
  So, comparing with Eqn(3.73) above, it can be seen that the y- and z-components of  Amp\`{e}re's law
  are just special cases of the general relation between x- and t-derivatives:
 \begin{equation}
 \left. -\frac{1}{c}\frac{\partial ~ }{\partial t}\right|_{x_q} = \left. \beta_u \frac{\partial ~}{\partial x_q}\right|_t
  \end{equation}
  that is a necessary consequence of the space-time functional dependence of both the RCED and CEM potentials
   and fields in the frame S. Thus the x- and y-components of  Amp\`{e}re's law are valid for both the RCED and CEM
   fields of a moving charge.
   \par Note that for both the CEM and the RCED fields the existence of the `displacement current' term
  $-(1/c)\partial \vec{E}/\partial t$ in Amp\`{e}re's law, originally `added by hand' for reasons of
     symmetry, by Maxwell, is a necessary consequence of the electric field divergence equation and the relation
     (4.28) between $x$- and $t$-derivative operators.
    \par Because of Eqn(4.15) and the universal validity of the Faraday-Lenz law, the derivation of the electromagnetic
    wave equation for the electric CEM field follows exactly the algebra of Eqn(4.10)-(4.11) above. By the same token
    the wave equation for the magnetic CEM field is obtained by combining the curl of (4.4) with the time derivative
    of (4.3) as for the classical fields $\vec{\E}$ and  $\vec{\B}$. However, this is no longer true for the RCED
    fields. Straightforward calculation using Eqns(4.14) and (4.23) and the partial derivatives given in
    Eqns(B4)-(B8) of Appendix B, shows that for the RCED fields the wave equations
     for the electric and magnetic fields are replaced by the following expressions:
  \par \underline{RCED fields}
 \begin{eqnarray}
    \vec{\nabla}^2 E_x & = & \frac{1}{c^2}\frac{\partial^2  E_x}{\partial t^2}+
   \frac{3 Q}{r^4}\left(\gamma_u-\frac{1}{\gamma_u}\right)(5\sin^2\psi-2)\cos \psi \\
    \vec{\nabla}^2 E_y & = & \frac{1}{c^2}\frac{\partial^2  E_y}{\partial t^2}+
   \frac{3 Q}{ r^4}\left(\gamma_u-\frac{1}{\gamma_u}\right)(5\sin^2\psi-4)\sin \psi \cos \phi \\
   \vec{\nabla}^2 E_z & = & \frac{1}{c^2}\frac{\partial^2  E_z}{\partial t^2}+
   \frac{3 Q}{ r^4}\left(\gamma_u-\frac{1}{\gamma_u}\right)(5\sin^2\psi-4)\sin \psi \sin \phi \\
  \vec{\nabla}^2 B_x & = & \frac{1}{c^2}\frac{\partial^2  B_x}{\partial t^2} \\
   \vec{\nabla}^2 B_y & = & \frac{1}{c^2}\frac{\partial^2  B_y}{\partial t^2}-
   \frac{3 \beta_u Q}{ r^4}\left(\gamma_u-\frac{1}{\gamma_u}\right)(5\sin^2\psi-4)\sin \psi \sin \phi \\
  \vec{\nabla}^2 B_z & = & \frac{1}{c^2}\frac{\partial^2  B_z}{\partial t^2}+
   \frac{3 \beta_u Q}{ r^4}\left(\gamma_u-\frac{1}{\gamma_u}\right)(5\sin^2\psi-4)\sin \psi \cos \phi
  \end{eqnarray}
  \par Summarising, the following results have been obtained in this section:
   \begin{itemize}
   \item[(a)] The force fields of both RCED and CEM, as well as the classical fields $\vec{\E}$ and  $\vec{\B}$,
     satisfy the the magnetic field divergence equation (4.2) and the Faraday-Lenz law (4.3), in any reference
     frame, due to 3-vector identities. The equations are then covariant for any fields defined by the equations
     (3.19) and (3.20) for an arbitary 4-vector potential $A$, and are independent of the space-time functional
      dependence of $A$ or of the fields.
   \item[(b)] The electric field divergence equation (4.1) is satisfied, in any inertial frame, for CEM but not
         for RCED. In the latter case the divergence is given by Eqns(4.14) and (4.16) for the frames S and S',
          and (4.1) is satisfied only in the frame S$^*$ where the source charge is at rest.  
    \item[(c)]  In both RCED and CEM the x-component of the left side of Amp\`{e}re's law may be derived from
       the electric field divergence equation by use of the relations (3.73) and (4.20). The resulting
    equation reduces to (4.4) for the CEM fields, but not for the RCED ones where, in the frame S, the 
    non-covariant result Eqn(4.23) is obtained instead.
      \item[(d)] In both RCED and CEM, the y- and z-components of  Amp\`{e}re's law for the fields of
     a uniformly moving charge are verified and shown to be a direct consequence of the universal relations
       (4.20) between transverse electric and magnetic field components and the general relation
       (3.73) or (4.28) between $x_q$- and $t$-derivatives. The latter follows from the space-time functional
     dependence of the RCED and CEM fields. This gives a natural explanation, in this case, of Maxwell's
     `displacement current' in  Amp\`{e}re's law (4.4) that was originally `added by hand'.   
     \item[(e)] Derivation of the wave equations for the electric and magnetic fields proceeeds, for the CEM fields,
          in a similar manner as for the classical fields  shown in Eqns(4.10)-(4.12), due to the validity
    of Eqn(4.1) in this case. This is no longer true for RCED  where $\vec{\nabla} \cdot \vec{E} \ne 0$ and where
     the wave equations are replaced by Eqns(4.30)-(4.35). Only $B_x$ (which vanishes for a uniformly
     moving source charge) satisfies a wave equation. 
    \end{itemize}    
  \par What now is the impact of these results on the status of the discovery of the Lorentz transformation
  by requiring covariance of the free-space Maxwell equations, and of Maxwell's derivation of `electromagnetic 
  waves' by combining three of his equations: (4.1), (4.3) and (4.4) for `waves in the electric field', and
 (4.2), (4.3) and (4.4) for `waves in the magnetic field'? 
    \par Firstly, it is clear that the derivation of the Lorentz transformation has been carried out under 
         the hypothesis that the electric and magnetic fields are classical, with the
     space-time functional dependence of (3.56). This is true for neither to RCED or CEM force fields 
     discussed so far in the present paper. The actual functional dependence is similar to that
     of Eqns(3.54) and (3.55). This implies that the space and time derivatives transform, not according 
     to Eqns(3.33)-(3.35), appropriate for a classical field, but rather according to:
  \begin{equation}
   \frac{\partial ~}{\partial x_q^*} = \frac{\partial ~}{\partial x_q} = \frac{\partial ~}{\partial x'_q}     
  \end{equation}
 and
  \begin{equation}
 -\frac{1}{c} \frac{\partial ~}{\partial \tilde{t}} = \tilde{\beta} \frac{\partial ~}{\partial \tilde{x}},~~~~~ 
   \tilde{t} = t,t';~~\tilde{x} = x_q, x'_q
  \end{equation}
     with appropriate values of $\tilde{\beta}$ in the frame S ( $\tilde{\beta}= \beta_u$)
     or S' ( $\tilde{\beta}= -\beta_w$). Indeed, as shown in Appendix A, the relation (4.36) is essential for
    the derivation of the CEM fields from the `present position' potentials in (3.51) and(3.52).
  \par So the Lorentz transformation has been derived by requiring covariance of hypothetical
   classical electric and magnetic fields with space-time functional dependence markedly
    different from the actual RCED or CEM force fields of classical electrodynamics.
    Does this mean that the Lorentz
    transformation is incorrect? The answer to this question is definitely `no', because the Lorentz
    transformation may be independently derived from other simple axioms that are quite unrelated to 
    classical electromagnetism as invoked by Larmor, Lorentz or Poincar\'{e} ---covariance of Maxwell's
    equations, or, in the case of Einstein, by the hypothesis of the constancy of the speed of light.
    This point will be further discussed below. Thus the discovery, by Larmor, of the Lorentz transformation
    by requiring covariance of the `free space' Maxwell equations was of an essentially heuristic nature.
    \par Concerning the derivation of the wave equation for the electric field, this does not necessarily 
    require the fields to have the classical space-time functionality of Eqn(2.56). Since the CEM force
    fields, which are not classical, do, as shown above, satisfy all four Maxwell equations,
    the derivation proceeds as in
     Eqns(4.10)-(4.12) for the classical fields. However the corresponding potentials (3.109) and (3.110)
     do not transform as a 4-vector and the corresponding electric and magnetic fields of (3.7) and (3.8)
     give rise to relativistic paradoxes~\cite{JackTP}, and may, for certain spatial configurations
     of moving charges, predict a vanishing electromagnetic induction
     effect in the case of a stationary coil and a uniformly moving magnet~\cite{JHF3}. They are also 
      inconsistent with the results of recent experiments that demonstrated the non-retarded character of 
      force fields~\cite{JAP1}.
  
    \par It will be demonstrated in the following section, that, in any case,
  the Li\'{e}nard-Wiechert potentials do not correctly
    represent solutions of the inhomogeneous d'Alembert equation from which they were supposedly derived. That is, the CEM fields
    are the result of a mathematical error and so actually have no physical relevance. 
    
\SECTION{\bf{Force fields and radiation fields: Effects of virtual or real photons}}
  The free-space `Maxwell-like' equations for the RCED force fields of a uniformly  moving charge obtained in the
   previous section are, in the Frame S:
  \begin{eqnarray}
 \vec{\nabla} \cdot \vec{E} & = &\frac{Q}{r^3}\left(\gamma_u-\frac{1}{\gamma_u}\right)(2-3\sin^2\psi) \\
   \vec{\nabla} \cdot \vec{B} & = & 0 \\
    \vec{\nabla} \times \vec{E} +\frac{1}{c}\frac{\partial\vec{B}}{\partial t} & = & 0 \\
 \vec{\nabla} \times \vec{B} -\frac{1}{c}\frac{\partial \vec{E}}{\partial t} & = & 
  \frac{Q \vec{\beta}_u}{r^3}\left(\gamma_u-\frac{1}{\gamma_u}\right)(2-3\sin^2\psi) 
  \end{eqnarray}
   The electric field divergence equation and the Amp\`{e}re law-like equation are not covariant, the covariance 
   breaking terms being of O($\beta_u^4$) in the first case and of O($\beta_u^5$) in the second. 
   \par Actually, however, all predictions concerning intercharge forces in classical electrodynamics are contained 
    in the single equation (1.2) that gives directly the force on a test charge for an arbitary position, velocity
    and acceleration of the source charge. These include: Coulomb's law, the Biot-Savart law, the Lorentz force
    equation and the machanical effects of the Faraday-Lenz law (5.3)\footnote{This is only true
      for applications of the Faraday-Lenz law to problems involving systems of charges or currents
       in uniform motion. Such effects are described, in QED, by the exchange of space-like virtual photons
   ~\cite{JHF1}. Applications involving accelerated charges, where real photons are radiated,
     require extra degrees of freedom to be included in the Lagrangian, and are not described by Eqn(1.2).}
     All of the equations (5.1)-(5.4) as well as
 the definitions (3.19) and (3.20) of electric and magnetic fields are derived from simple mathematical
   substitutions en route to derive (1.2) from the postulates described in the Introduction. Thus Eqns(5.1)-(5.4)
   contain no information on inter-charge forces not already contained in Eqn(1.2).  For example. (5.2) and 
   (5.3), which are the same as the corresponding Maxwell equations (4.2) and (4.3), follow immediately as 3-vector
   identities from the definitions of the electric and magnetic fields. Note also that Eqn(5.4) contains Maxwell's
   famous and mysterious `displacement current' term $\partial \vec{E}/ c \partial t$ which is a direct consequence
   of (5.1) given given (3.73) and (4.20). In fact (5.4) may be derived from (5.1) ---it is not an independent
   equation. This conclusion is equally valid for the CEM force fields.
   \par When $u \rightarrow  0$,  $\gamma_u \rightarrow  1$ Eqn(5.1) reduces to the usual electric field divergence
   equation (4.1). Since Coulomb's law is the basic physical postulate of RCED, there is no difference between CEM and
   RCED as far as electrostatics is concerned. The same is true for magnetostatics when only forces between neutral
   conductors carrying steady currents are considered and corrections of O($\beta^2$) and higher are neglected.
   The new feature of RCED is that nominally neutral conductors will, according to Eqn(3.3), have a small transverse
   electric field corresponding to an effective charge $\simeq -|e|\beta_u^2$ for each conduction electron where
   $e$ is the electron charge, and a longitudinal electric field corresponding to an effective charge $\simeq +|e|\beta_u^2$
   for each conduction electron. This is because the magnitudes of the effective source charges for transverse (longitudinal)
   fields are greater (less) than those of the positive ions (assumed to be essentially at rest) in the bulk matter of
   the conductor. On averaging over  an isotropic distribution
     of directions, for a source of fixed speed, the electric  field is increased, at leading order in
     $\beta_u$, by the fraction $\beta_u^2/6$. Experimental evidence for a tranverse field
     $\simeq \beta_u^2$ in the neighbourhood of a supercondicting coil was already published some three decades 
     ago~\cite{Edwards}. It is interesting to note that the CEM 
     formula (3.7) predicts the same transverse field as the RCED one (3.5) when $\psi = \pi/2$, but 
     a two-times larger effective charge for a longitudinal field when $\psi = 0$ due to the
  $(\gamma_u)^{-2}$ dependence of such a field. 
    \par It is important to stress that, since the force due to an electric field on a test charge does not
     depend on the velocity of the latter, these small residual electric fields in magnetostatics, present
      in both CEM and RCED, have no
    mechanical effect on neutral systems. In particular, the very stringent limits on the neutrality of atoms or
     molecules~\cite{JGK,ZCVWH},
    by detecting any force to which they are subjected in strong electric fields, are not affected by the existence of these fields.
     \par In summary, all the purely mechanical effects of classical electromagnetism ---including relativistic corrections
       to all orders in $\beta$,
      are given by the single equation (1.2) which contains information solely on the relative spatial positions and
     velocities of the interacting charges ---no fields. As discussed at length in Reference~\cite{JHF1}, according 
     to quantum electrodynamics, these forces are produced by the exchange of space-like virtual photons between the 
     charges. The corresponding interaction is instantaeous ---not retarded--- in the centre of mass frame of the charges.
     However these phenomena are, by far, not the whole content of classical or quantum electrodynamics. It remains to
    describe radiative processes in which, according to quantum electrodynamics, real photons are created, propagated
     and destroyed. It is in this sector that both the utility of Maxwell's equations and the physical meaning of the fields or
    potentials that they contain will become evident, as well as any shortcomings in their application. Contrary to the
     description of mechanical effects, fully
    described by Eqn(1.2), for which the field concept is inessential, the route followed for the classical description
    of radiative processes will be the well-trodden one of 19th Century physics ---identification of a suitable
    partial differential equation to describe a physical system, solution of the equation subject to certain
    boundary conditions and the physical interpretation of the solutions found. For this programme, the concepts
    of potentials and fields (although not necessarily classical ones in the sense of Eqn(3.56)) are of
    fundamental importance.
    \par To distinguish the fields and potentials in the following discussion from the force fields and associated
    potentials considered hitherto, roman characters will be used. The analysis is based on the following
    definitions and differential equations, which are first presented in full, and subsequently discussed:
    \begin{eqnarray}
   \vec{\Ei} & \equiv & -\vec{\nabla}\Ai_0 -\frac{1}{c}\frac{\partial \vec{\Ai}}{\partial t} \\
   \vec{\Bi} & \equiv & \vec{\nabla} \times \vec{\Ai}  \\
     \rho^*(\vec{x}_J^*)& \equiv & \frac{1}{V_R(\vec{x}_J^*)} \sum_{i \subset R} Q_i \\
     \Ji(\vec{x}_J,t) & \equiv & (  \gamma_u \rho^*; \gamma_u \vec{\beta}_u \rho^*) = \frac{u  \rho^*}{c} \\
     \vec{\nabla} \cdot \vec{\Ai} & + & \frac{1}{c}\frac{\partial \Ai_0}{\partial t}  = 0 \\
      \vec{\nabla} \cdot \vec{\Ei} & = & 4 \pi \Ji_0 \\
       \vec{\nabla} \times \vec{\Bi} & - &\frac{1}{c}
      \frac{\partial \vec{\Ei}}{\partial t} =  4 \pi \vec{\Ji} 
    \end{eqnarray}
      All fields, potentials and currents are defined in the frame S where the source and test charges are in motion,
      whereas the electric charge density in Eqn(5.7) is defined in the frame S$^*$ in which the source 
      charges $Q_i$ are at rest. $V_R(\vec{x}_J^*)$ is the volume of a small  region, $R$, at
      $\vec{x}_J^*$ containing the discrete charges $Q_i$.  Eqns(5.5) and (5.6) are the conventional covariant
      definitions of electric and magnetic fields in terms of the 4-vector potential, while (5.8) defines
     the 4-vector current density, $\Ji$. The 4-vector potential, $\Ai$, will be obtained as a solution of
     the Maxwell equations with sources, (5.10) and (5.11), and is also assumed, initially, to satisfy the Lorenz
     condition (5.9). The latter is conventionally  presented as a particular choice of gauge (Lorenz gauge
      ~\cite{JackOkun}) for the definition of  $\Ai$. However, as shown in ~\cite{JHF1},
     for the RCED force field potentials in (3.49) and (3.50), this formula is an identity following from the
      relation (4.36) connecting spatial and temporal partial derivatives. It may be noted that the magnetic field 
    divergence equation and the Faraday-Lenz law, that follow as 3-vector identities from (5.6) and (5.5)
    respectively, do not appear in the following discussion. This may be contrasted with the discussion
    of `electromagnetic waves' based on the free-space Maxwell equations in Section 4. This is indicative
   of the heuristic, rather than fundamental, nature of both derivations.  
    \par As is well known~\cite{JackDE}, after expressing (5.10) and (5.11) in terms of $\Ai$ with the aid 
    of (5.5) and (5.6), the Lorenz condition (5.9) may be used to eliminate $\vec{\Ai}$ in (5.10) or
     $\Ai_0$ in (5.11) so as to yield the inhomogeneous d'Alembert equations:
        \begin{eqnarray}
      \nabla^2 \Ai_0 -\frac{1}{c^2}\frac{\partial^2 \Ai_0}{\partial t^2} & = & -4 \pi \Ji_0  \\
       \nabla^2 \vec{\Ai} -\frac{1}{c^2}\frac{\partial^2  \vec{\Ai}}{\partial t^2} & = & -4 \pi \vec{\Ji}
       \end{eqnarray}
     The particular solutions, in the frame S, of these equations are~\cite{JackGF}:
     \begin{equation}
     \Ai_{\mu}(\vec{x}_q,t)  = \int dt' \int d^3 x_J(t') \frac{\Ji_{\mu}(\vec{x}_J(t'), t')}
       {|\vec{x}_q -\vec{x}_J(t')|} \delta(t'+\frac{|\vec{x}_q -\vec{x}_J(t')|}{c}-t)
     \end{equation} 
        The delta-function in (5.14) ensures that all quantities on the right side of the equation are evaluated
      at the retarded time $t'$ given by
  \begin{equation}
     t' = t - \frac{|\vec{x}_q -\vec{x}_J(t')|}{c} \equiv t - \frac{r_J}{c}  
     \end{equation}
      As in the previous discussion, the particular case of a single source charge will now be considered.
      For such as charge, situated at $\vec{x}_Q(t')$, the current density in (5.8) becomes
  \begin{equation}
    \Ji_Q(\vec{x}_J(t'), t') = \frac{ Q u}{c} \delta (\vec{x}_J(t')-\vec{x}_Q(t')) 
     \end{equation}
    Substituting (5.16) in (5.14) and integrating over $\vec{x}_J$ gives \footnote{The notation specifying
    the space-time point at which the potential is evaluated $\Ai_{\mu}(\vec{x}_q,t)$
    does not imply that $\Ai_{\mu}$ is a `classical' field in the sense of Eqn(3.56) where $x_q$,  $y_q$, $z_q$ and $t$
   are independent variables. In virtue of the definition of $r$ in (5.18), $x_q$ and $t$ are not independent
    variables. c.f. Eqns(C.7), (C.8) and (C.15) of Appendix C.}  
     \begin{equation}
     \Ai_{\mu}(\vec{x}_q,t)  = \frac{Q u_{\mu}}{c} \int dt'  \frac{\delta(t'-t'_Q)}
       {|\vec{x}_q -\vec{x}_Q(t')|}
     \end{equation} 
   where 
  \begin{equation}
     t'_Q = t - \frac{|\vec{x}_q -\vec{x}_Q(t'_Q)|}{c} \equiv t - \frac{r}{c}  
  \end{equation}
 The components of the 4-vector $\Ai_{\mu}$ are then given by (5.17) and (5.18) as:
   \begin{equation}
  (\Ai_0;\vec{\Ai}) = \left( \left. \frac{Q \gamma_u}{r} \right|_{t' = t'_Q};
     \left. \frac{Q \gamma_u \vec{\beta}_u}{r}\right|_{t' = t'_Q} \right)
   \end{equation}
   Subsituting (5.19) into (5.5) and (5.6) and evaluating the derivatives as shown in Appendix C 
   gives the following results for the electric and magnetic fields corresponding to $\Ai_{\mu}(\vec{x}_q,t)$:
     \begin{eqnarray}
   \vec{\Ei}(\vec{x}_q,t) & = & \left. \frac{Q \gamma_u}{1-\hat{r}\cdot \vec{\beta}_u}\left[ 
      \frac{[\hat{r}-\vec{\beta}_u (\hat{r}\cdot \vec{\beta}_u)]}{r^2}
    +  \frac{[ \gamma_u^2 \beta_u \dot{\beta}_u (\hat{r}- \vec{\beta}_u)- \dot{\vec{\beta}_u}]}{cr}\right] \right|_{t' = t'_Q} \\
  \vec{\Bi}(\vec{x}_q,t) & = & \left. \frac{Q \gamma_u (\vec{\beta}_u \times \hat{r})}{1-\hat{r}\cdot \vec{\beta}_u}\left[ 
      \frac{1}{r^2}
    + \frac{\gamma_u^2 \dot{\beta}_u}{ \beta_u c r}\right] \right|_{t' = t'_Q} 
  \end{eqnarray}
   The dotted symbols in these equations denote derivatives with respect to $t'$ and $\hat{r} \equiv \vec{r}/r$.
   The first terms on the right sides of (5.20) and (5.21), with dependence $1/r^2$, are retarded force fields.
   Note that, apart from the common factor $1/(1-\hat{r}\cdot \vec{\beta}_u)$ and the retarded time argument
   these are the same as the RCED force fields (3.5) and (3.6) for the case of a uniformly
   moving source charge ($\dot{\beta}_u =  \dot{\vec{\beta}_u} = 0$). They are quite different from the CEM force fields
   with retarded time arguments~\cite{JackLW,LLLW} derived from the retarded Li\'{e}nard-Wiechert potentials~\cite{PPFMC}
    of a point charge:
      \begin{eqnarray}
   A_0(LW) & = & \left. \frac{Q}{r(1-\hat{r}\cdot \vec{\beta}_u)}\right|_{t' = t'_Q}  \\
   \vec{A}(LW) & = & \left. \frac{Q \vec{\beta}_u}{r(1-\hat{r}\cdot \vec{\beta}_u)}\right|_{t' = t'_Q} 
  \end{eqnarray}
   The reason for this difference will be discussed at the end of the present section.
  \par As is well known, the second terms on the right sides of (5.20) and (5.21) with $1/r$ dependence,
   each proportional to the acceleration of the source charge, give the classical description of radiative 
   procesess, that is the production and propagation of real photons. Equation (5.18) predicts that these real
   photons propagate in free space at the speed $c$. It will be shown below that the corresponding
   potential and fields $\Ai_{rad}$, $\Ei_{rad}$ and $\Bi_{rad}$ are closely related (and, in fact, equivalent to)
   the quantum mechanical amplitude for the production of one or more real photons by the accelerated source
   charge.
    \par As pointed out Reference~\cite{JHF1}, QED predicts that the force fields, associated uniquely with the 
    exchange of space-like virtual photons, are instantaneous, so that the force fields in (5.20) and (5.21), 
    corresponding to photons propagating at the speed $c$, are not consistent with this requirement of QED or
    the known kinematical properties of space-like virtual photons. They are also inconsistent with experiment
    ~\cite{JAP1}.
    \par For the further consideration of the radiation fields it will be assumed, for definitness, that
      the source charge undergoes simple harmonic motion
        of frequency $\omega/(2\pi)$ parallel to the x-axis, so that it constitutes
     a system of fixed orientation with a harmonically varying dipole moment:
     \begin{equation}
      \vec{p}(t'_Q) =  Q \vec{x}_Q(t') = \vec{p}_0 e^{-i \omega t'_Q}
     \end{equation}
   It follows from (5.24) that:
     \begin{equation}
     \frac{Q  \dot{\vec{\beta}_u}}{c} = -k^2 \vec{p}
     \end{equation}
    where $k \equiv \omega/c \equiv 2 \pi/\lambda$, where $\lambda$ is the `wavelength' of the radiated photons.
    Subsituting (5.25) in (5.20) and (5.21) gives the following expressions for the radiation fields:
    \begin{eqnarray}
     \vec{\Ei}_{rad}(\vec{x}_q,t) &  =  & \left[ \frac{k^2 \gamma_u^3}{r(1-\hat{r}\cdot \vec{\beta}_u)}     
       (\vec{p}-\beta_u  p \hat{r}) \right]_{t' = t'_Q} \\
  \vec{\Bi}_{rad}(\vec{x}_q,t) &  =  & \left[ -\frac{k^2 \gamma_u^3 
    (\vec{p} \times \hat{r})}{r(1-\hat{r}\cdot \vec{\beta}_u)}     
     \right]_{t' = t'_Q}
     \end{eqnarray}
   where $\vec{p} = \hat{\imath} p$.
   In the non-relativistic limit, $\beta_u \ll 1$, these equations simplify to:
    \begin{eqnarray}
     \vec{\Ei}_{rad}(\vec{x}_q,t) &  =  &  \frac{k^2\vec{p}_0 e^{-i\omega t'_Q}}{r} \\
  \vec{\Bi}_{rad}(\vec{x}_q,t) &  =  & -\frac{k^2(\vec{p_0} \times \hat{r})}{r} e^{-i\omega t'_Q}      
     \end{eqnarray}
    The spatial amplitude of the harmonic oscillations is given by
      $|x_Q| = |\beta_u|/k = \lambda|\beta_u|/(2 \pi)$. Imposing the `far zone' condition $r \gg |x_Q|$
     then implies that 
  \begin{equation}
  \frac{r}{\lambda} \gg \frac{|x_Q|}{\lambda} = \frac{|\beta_u|}{2 \pi}
    \end{equation}
    If, in addition, $|x_Q| \ll \lambda$, the non-relativistic limit condition $\beta_u \ll 1$
    is automatically satisfied. In the far zone the $t'$ dependence of $r$ in (5.28) and (5.29)
    is negligible. The condition (5.30) together with $|x_Q| \ll \lambda$ prescribe the conditions 
    for `dipole radiation' where the produced real photons are observed in the far zone.
   \par The time-averaged photon energy flux $dP$ radiated into solid angle $d\Omega$ in the direction
    $\hat{r}$ is given by the Poynting vector as~\cite{JackPV,SchwartzPV}\footnote{In Eqn(5.31) the star superscript
     on $ \vec{\Bi}_{rad}$ denotes complex conjugation, not the frame.}:
   \begin{eqnarray}
    \frac{dP}{d\Omega} &  = & \frac{c}{8 \pi}{\cal R}e[ r^2 \hat{r}
     \cdot (  \vec{\Ei}_{rad} \times \vec{\Bi}_{rad}^*)] =  \frac{c k^4}{8 \pi}
    [\hat{r} \cdot (\vec{p}_0 \times (\hat{r} \times \vec{p}_0))] \nonumber \\
       & = &  \frac{c k^4}{8 \pi} [|p_0|^2 -|p_0 \cdot \hat{r}|^2] =  \frac{c k^4}{8 \pi}|p_0|^2 \sin^2 \psi
    \end{eqnarray}
     This is the usual formula~\cite{JackDR} for the energy flux per unit solid angle in dipole radiation.
     Note however that the electric radiation field (5.28) differs from the conventional one in that
     it is {\it not} transverse to the direction $\hat{r}$ but is instead parallel to the 
     direction of the oscillating dipole moment. The magnetic field in (5.29) is however perpendicular
     to both $\hat{r}$ and $\vec{p}$ as in the conventional treatment.
     \par At sufficiently large distances from the dipole source only one or zero photons
      are expected to be observed in a spatial volume $dV =  dS dr$, where 
     $dS$ is an area element perpendicular to $\vec{r}$. The probability
    to observe two or more photons, though non-vanishing, is then much less than that to observe only
    one. As already pointed out in Ref.~\cite{JHF5}, in these circumstances, `inverse correspondence'
    gives a connection between the classical and quantum mechanical descriptions of radiative processes
    involving real photons. In particular, Eqn(5.31) may be directly related to the quantum mechanical
    description of dipole radiation. Since $d\Omega = r^2 dS$, Eqn(5.31) may be written as:
     \begin{equation}
     \frac{dP}{dS} = \frac{c}{ 8 \pi} \frac{ k^4 |p_0|^2 \sin^2 \psi}{r^2} = c \bar{\rho}_{\gamma}(\vec{r}) E_{\gamma}
    \end{equation}
      where $\bar{\rho}_{\gamma}(\vec{r})$ is the time averaged number density of photons, each with energy $E_{\gamma}$.
      Thus
        \begin{equation}
   \bar{\rho}_{\gamma}(\vec{r}) = \frac{ k^4 |p_0|^2 \sin^2 \psi}{8 \pi r^2  E_{\gamma}}
      = \frac{|B_{rad}|^2}{8  \pi E_{\gamma}}
    \end{equation} 
     The equality of the first and last members of (5.33) was previously noted,~\cite{JHF5}, for the
     case of a plane electromagnetic wave. For sufficiently large values of $r$,  $\bar{\rho}_{\gamma}(\vec{r}) dV$
    becomes, to a very good approximation, the probability that a photon is observed in the volume
    element $dV$ at $\vec{r}$ at any instant for which the backward light cone of the field point at the source
   overlaps the latter
    during its phase of continuous oscillation. In this case $\bar{\rho}_{\gamma}(\vec{r}) = |\Psi(\vec{r})^2|$ 
  where $\Psi(\vec{r})$ is the quantum probability amplitude (conventionally called a `wavefunction')
   for detection of a photon at $\vec{r}$. Indeed, even the detailed time structure of the photon
   detection events is already implicit in the classical calculation. 
   \par Taking the square root of (5.33), inserting the harmonic time dependence of $B_{rad}$
   from Eqn(5.29), setting $t'_Q = t_{\gamma}$, and using (5.18) gives, for the quantum probability amplitude:
   \begin{eqnarray}
   \Psi(\vec{r},t,t_0) & = &\frac{\Mg}{r} \exp [-i \omega(t_{\gamma}-t_0)] \nonumber \\
          & = & \frac{\exp[-i(\omega(t-t_{\gamma})-k r)]}{r} \Mg  \exp[-i \omega(t -t_0 -\frac{r}{c})] 
    \nonumber \\
       & = & \langle \vec{x}_q,t|\gamma| \vec{x}_Q,t_{\gamma}\rangle \Mg \langle  t_{\gamma}|Q|t_0 \rangle
    \end{eqnarray}
     where
        \begin{equation}
     \Mg \equiv \frac{k^2 |p_0|\sin \psi}{\sqrt{8 \pi E_{\gamma}}}
      \end{equation}
    The last line of Eqn(5.34) has the structure of a time-ordered Feynman path amplitude ~\cite{JHF6}.
    The rightmost amplitude in (5.34) is the temporal propagator of the oscillating source charge from 
   the time $t_0$, at which harmonic motion starts, to the time, $t_{\gamma}$, at which the detected 
   photon is created. The time-independent amplitude $\Mg$ is that for photon creation, while the first
   amplitude on the right side of (5.34) is the space-time propagator of the photon between its
   production and detection points. Note that since $r/(t-t_{\gamma}) = c = \omega/k$ the argument of
    the exponential in the photon propagator vanishes, so that the latter is simply $1/r$. It can be seen
  that the phase of the probability amplitude, of no relevance for calculation of the photon flux, but
    essential for the description of quantum interference phenomena, is uniquely that associated with
    the source charge~\cite{FeynQED,JHF6}. The probabilty amplitude of (5.34) may be multiplied by the factor 
      $\exp[-i\phi_0]$ without affecting any physical prediction. The constant real phase $\phi_0$
    corresponds in the clasical calculation to the phase of the oscillator when $t = t_0$.
     \par It is instructive to compare Eqn(5.33) with the similar formula describing photon production
    by atomic dipole radiation~\cite{SakuraiDR}:
        \begin{equation}
   \bar{\rho}_{\gamma}(\vec{r}) = \frac{ k^4 |\langle B |p_{BA}| A\rangle|^2 \sin^2 \psi}{8 \pi^2 r^2  E_{\gamma}}
    \end{equation}   
   This quantum mechanical formula is obtained from the classical formula (5.33) by the replacement
  \[ p_0 \rightarrow \frac{\langle B |p_{BA}| A\rangle}{\sqrt{\pi}} \]
   where $p_{AB}$ is the quantum mechanical electric dipole operator associated with the transition
    between the atomic states $A$ and $B$. The probabilty amplitude corresponding to (5.34) for the case 
    of spontaneous decay of an excited atom can be found in Ref~\cite{JHF6}, it is:
  \begin{equation}
 \Psi(\vec{r},t,t_0) = \frac{{\cal M}_0}{r}
   \exp[-i(\omega-\frac{1}{2 \tau_A})(t -t_0 -\frac{r}{c})]
    \end{equation}
   Here $t_0$ is the time of production of the excited atom in the state $A$ and $\tau_A$ is the mean
    lifetime of the state $A$. In this case:
 \begin{equation}
      E_A - E_B = E_{\gamma} = \hbar \omega
 \end{equation}
      and, from Eqn(5.36): 
 \begin{equation}
  {\cal M}_0 = \frac{E_{\gamma}^{\frac{3}{2}}\langle B |p_{BA}| A\rangle \sin \psi}{\sqrt{8} \pi \hbar^2 c^2}
 \end{equation}
  \par In deriving the d'Alembert equations (5.12) and (5.13), of which (5.19) is the solution, the Lorenz 
  condition (5.9) was assumed. It is of interest to check the {\it a posteriori} validity of (5.9) by 
   direct substitution in it of the solution (5.19). It is shown in Appendix C that this gives, for the two
   terms on the left side of the Lorenz condition:
    \begin{eqnarray}
   \vec{\nabla} \cdot \vec{\Ai} & = & \left.-\frac{Q \gamma_u (\hat{r} \cdot \vec{\beta}_u)}
   {1-\hat{r} \cdot \vec{\beta}_u}
  \left[\frac{1}{r^2}+ \frac{\gamma_u^2 \dot{\beta}_u }{c \beta_u r}\right] \right|_{t' = t'_Q} \\
   \frac{1}{c}\frac{\partial \Ai_0}{\partial t}  & = & \left. \frac{Q \gamma_u}{1-\hat{r} \cdot \vec{\beta}_u}
  \left(\frac{\hat{r}\cdot \vec{\beta}_u}{r^2}
   + \frac{\gamma_u^2 \beta_u \dot{\beta}_u}{cr}\right)\right|_{t' = t'_Q}
  \end{eqnarray}
    Adding (5.40) and (5.41):
   \begin{equation}
  \vec{\nabla} \cdot \vec{\Ai} +  \frac{1}{c}\frac{\partial \Ai_0}{\partial t} =
   \left. \frac{Q \gamma_u \dot{\beta}_u }{c r(1-\hat{r} \cdot \vec{\beta}_u)}\left[
 \beta_u-\frac{\hat{r}\cdot \vec{\beta}_u}{\beta_u}\right]\right|_{t' = t'_Q }
   \end{equation}
    Thus, the terms in the Lorenz condition associated with a constant source velocity verify it as an identity
   in the same way as the instantaneous potentials (3.104), (3.105) corresponding to the RCED force fields of
   a uniformly moving charge. However, the terms containing derivatives of the source velocity, and so associated
     with the radiative fields, do not respect the condition. Thus, although (5.9) was assumed
   to be valid to derive (5.12) and (5.13), the Lorenz condition holds only for the 
    `force fields', $ \propto 1/r^2$ not the `radiative' ones $ \propto 1/r$.
     For simple harmonic motion of the source according 
    to (5.24), and taking the non-relativistic limit $\beta_u \rightarrow 0$, (5.42) simplifies to:
 \begin{equation}
  \vec{\nabla} \cdot \vec{\Ai} +    \frac{1}{c}\frac{\partial \Ai_0}{\partial t} = \frac{k^2 p}{r} 
   \end{equation}       
 In this limit the term on the right side of this equation comes only from the $\vec{\nabla} \cdot \vec{\Ai}$ 
   term on the left side. 
   \par Since (5.19) and (5.24) give, in the non-relativistic limit:
  \begin{equation}
   \vec{\Ai} = \hat{\imath} \Ai = -\frac{i k \vec{p}}{r} = -\frac{i \hat{\imath} k p}{r} 
  \end{equation}
  it follows from (5.43) and (5.44) that that:
  \begin{equation}
  \vec{\nabla} \cdot \vec{\Ai} = i k \Ai = i |\vec{k}| |\vec{\Ai}|
 \end{equation}
  It is common in text books on quantum mechanics or quantum field theory to associate
  $\vec{\Ai}$ with the polarisation vector, $\vec{\epsilon}$ of a free photon with
  wave number $\vec{k}$~\cite{HMPP,SakuraiPP}:
 \begin{equation}
  \vec{\Ai} \simeq \vec{\epsilon} \exp[i(\vec{k} \cdot \vec{x}-\omega t)]
 \end{equation}
 This equation gives:
  \begin{equation}
  \vec{\nabla} \cdot \vec{\Ai} \simeq 
    i \vec{k} \cdot  \vec{\epsilon} \exp[i(\vec{k} \cdot \vec{x}-\omega t)]
 \end{equation}
   Conventionally it is assumed that Gauge Invariance
   allows the choice $ \vec{\nabla} \cdot \vec{\Ai} =0$ (`Coulomb gauge'). It then follows from (5.47) that $\vec{k} \cdot  \vec{\epsilon} = 0$ so that
   the polarisation vector is transverse to the direction of propagation $\hat{k}$ of the photon. In fact (5.45)-(5.47)
    give instead:
    \begin{equation}
     \vec{k} \cdot \vec{\epsilon} = |\vec{k}| |\vec{\epsilon}|
  \end{equation}
  i.e. $\vec{\epsilon}$ is {\it parallel}  to  $\vec{k}$ if the ansatz (5.46) is assumed.
  Two remarks are in order:
  \begin{itemize}
   \item[(i)] In the non-relativistic limit discussed above, Eqn(5.43) shows that 
      $ \vec{\nabla} \cdot \vec{\Ai} \ne 0$, 
    so that the RCED calculation is not consistent with Coulomb gauge  $ \vec{\nabla} \cdot \vec{\Ai} =0$.

   \item[(ii)]  The phase of $\vec{\Ai}$ originates in the propagator of the photon
    source --it is not that of the `wavefunction of the photon' as the identification
    (5.46) suggests. If the `polarisation vector of the photon' is to be associated with $\vec{\Ai}$ then it is neither
     perpendicular to $\hat{k}$ nor parallel to it, but as is clear from (5.44), parallel to the direction of the
     dipole source. Thus neither the usual identification of $\vec{\Ai}$ with the `photon wavefunction' according
     to Eqn(5.46) nor the assumed validity of the condition $ \vec{\nabla} \cdot \vec{\Ai} =0$ 
     are tenable hypotheses in RCED.
  \end{itemize}
    
    \par In order to now discuss the space-time transformation properties of the radiation
   fields it is convenient to associate the frame S$^*$ with the one in which the average velocity, 
   due to its simple harmonic motion, of the oscillating charge, vanishes.The frame S then moves with constant
    velocity $c\vec{\beta}_u$ along the x-axis. Assuming the geometry of Fig.1, the radiation fields of
    (5.26) and (5.27) can then be written, in the frames S and  S$^*$, as:     
    \begin{eqnarray}
     \vec{\Ei}_{rad}(\vec{x}_q,t) & =  & \frac{k^2 \gamma_u^3 p}{r}\left[\hat{\imath}
    - \hat{\jmath}\frac{\beta_u \sin \psi}{1-\beta_u \cos \psi}\right] \\
     \vec{\Bi}_{rad}(\vec{x}_q,t) & =  & -\frac{ \hat{k} k^2 \gamma_u^3 p\sin \psi }{r(1-\beta_u \cos \psi)} \\
    \vec{\Ei}_{rad}^*(\vec{x}_q,t) & =  & \hat{\imath}\frac{(k^*)^2 p^*}{r} \\
    \vec{\Bi}_{rad}^*(\vec{x}_q,t) & =  & -\frac{\hat{k}(k^*)^2 p^* \sin \psi^* }{r}
   \end{eqnarray}
      It is assumed in these equations that:
   \begin{equation}
    \beta_u \gg \frac{kp}{Q}
   \end{equation}
   so that the velocity of the charge due to its simple harmonic motion has been 
   neglected as compared to $c\beta_u$
   \par If (5.49) and (5.50) are substituted  in (5.31), in order to calculate the photon energy flux
    per unit solid angle in the frame S, it is found that:
      \begin{equation}
    \frac{d P}{d \Omega} = \frac{c k^4 \gamma_u^6 |p_0|^2 \sin^2 \psi}{8 \pi (1-\beta_u \cos \psi)^2}
     \end{equation}
   which may be compared to the result (5.30) previously calculated for the frame S$^*$:
      \begin{equation}
    \frac{d P^*}{d \Omega^*} = \frac{c (k^*)^4  |p_0|^2 \sin^2 \psi^*}{8 \pi}
     \end{equation} 

    Eqn(5.54) may be checked by direct transformation of the differential energy flux (5.55) in the frame S$^*$
    into the frame S. Noting that $dP^* = dn_{\gamma}E^*_{\gamma}$ where $dn_{\gamma}$ is the photon flux in the
    solid angle element $d \Omega^*$ and $E^*_{\gamma}$ is the photon energy, and assuming that the flux of photons
   is conserved in the transformation from the frame S$^*$ to the frame S., the energy flux in the frame S
    is given by (5.55) as:
     \begin{equation}
    \frac{d P}{d \Omega} = \frac{c k^4 |p_0|^2 \sin^2 \psi}{8 \pi}\left(\frac{k^*}{k}\right)^4\left(\frac{\sin \psi^*}
     {\sin \psi}\right)^2 \left(\frac{E_{\gamma}}{E_{\gamma}^*}\right) \frac{d \Omega^*}{d \Omega}
     \end{equation}
    The oscillating electric dipole, as  viewed in the frame S, constitutes a moving clock whose observed frequency
  is slowed down by the relativistic time dilatation effect. In consequence $p = p_0 \exp(-i\omega t') =
    p_0 \exp(-i\omega^* t'/\gamma_u)$ whereas  $p^* = p_0 \exp(-i\omega^* t^{*})$. It follows that:
      \begin{equation}   
     k = \frac{\omega}{c} =  \frac{\omega^*}{\gamma_u c} = \frac{k^*}{\gamma_u}
  \end{equation} 
     The kinematics of massless photons gives the relations:
       \begin{equation}
      \frac{\sin \psi^*}{\sin \psi} = \frac{E_{\gamma}}{E_\gamma^*} = \frac{1}{\gamma_u(1-\beta_u \cos \psi)}
      \end{equation}
     and 
      \begin{equation}
      \cos \psi^* = \frac{ \cos \psi - \beta_u}{1-\beta_u \cos \psi}
      \end{equation}
    Differentiating (5.59) gives:
        \begin{equation}
     \frac{d \Omega^*}{d \Omega} = 
      \frac{d \cos \psi^*}{d \cos \psi} = \frac{1} {\gamma_u^2 (1-\beta_u \cos \psi)^2} 
    \end{equation}  
     Combining (5.56)-(5.60) gives:
          \begin{equation}
    \frac{d P}{d \Omega} = \frac{c k^4 |p_0|^2 \sin^2 \psi}{8 \pi \gamma_u(1-\beta_u \cos \psi)^5}
     \end{equation}
      which is markedly different from the prediction (5.54) given by the Poynting vector of the
      radiation fields in the frame S.
   Evidently the formula (5.31) relating the 
   photon energy flux to the Poynting vector is only valid, as a non-relativistic approximation, in the
   frame S$^*$, where good agreement is found with the analogous quantum mechanical predictions in (5.34) or (5.36).
    \par The incompatiblity between the solution (5.19) of the d'Alembert equations (5.12) and (5.13) with the
    retarded Li\'{e}nard-Wiechert (LW) potentials (5.22) and (5.23) will now be discussed. Inverting the order
    of the space and time integrations in (5.14)
  \footnote{In Section 6.6 of Ref.~\cite{Jack1}, describing Green function solutions of
    the inhomogeneous d'Alembert Equation, the temporal and spatial integrals are nested
   as in Eqn(5.14) above. Eleswhere in the book, where radiative effects 
    were discussed (Eqn(9.2) of Section 9.1) the order was inverted to correspond to that of Eqn(5.62)
    above. This ordering was also implicit in the discussion of radiation from moving
    charges in Chapter 14, where, in a derivation of the LW potentials using a covariant relativistic
    notation, the $\delta$-function transformation (5.63) was employed to obtain the 
    retarded potentials (5.22) and (5.23) above.}, it may be written as:
     \begin{equation}
     \Ai_{\mu}(\vec{x}_q,t)  = \int d^3 x_J(t') \int dt' \frac{\Ji_{\mu}(\vec{x}_J(t'), t')}
       {|\vec{x}_q -\vec{x}_J(t')|} \delta(t'+\frac{|\vec{x}_q -\vec{x}_J(t')|}{c}-t)
     \end{equation} 
    The property of the Dirac $\delta$-function:
   \begin{equation}
     \delta [f(x)] = \frac{\delta(x-x_0)}{~~~~\left|\frac{\partial f(x)}{\partial x}\right|_{x= x_0}}
     \end{equation} 
  where $x = x_0$ is the solution of the equation $f(x) = 0$ may be used to rewrite (5.62) as:
     \begin{equation}
     \Ai_{\mu}(\vec{x}_q,t)  = \int d^3 x_J(t') \int dt' \frac{\Ji_{\mu}(\vec{x}_J(t'), t')}
       {|\vec{x}_q -\vec{x}_J(t')|(1-\hat{r}_J \cdot \vec{\beta}_J)} \delta(t'- t_J')   
     \end{equation} 
     where
     \begin{equation}
     \hat{r}_J = \frac{\vec{x}_q -\vec{x}_J(t_J')}{|\vec{x}_q -\vec{x}_J(t_J')|},
      ~~~~\vec{\beta}_J = \frac{1}{c}\frac{d \vec{x}_J}{d t'}
     \end{equation} 
 and
      \begin{equation}    
      t_J' = t - \frac{|\vec{x}_q -\vec{x}_J(t_J')|}{c}
   \end{equation} 
    Performing the integral over $t'$ gives:
      \begin{equation}
     \Ai_{\mu}(\vec{x}_q,t)  = \int d^3 x_J(t_J') \frac{\Ji_{\mu}(\vec{x}_J(t_J'), t_J')}
       {|\vec{x}_q -\vec{x}_J(t_J')|(1-\hat{r}_J \cdot \vec{\beta}_J)}
     \end{equation} 
       Specialising to the case of a point charge by substituting in (5.67) the current
     density of (5.16) then gives:     
       \begin{equation}
     \Ai_{\mu}(\vec{x}_q,t)  =\left. \frac{Q u_{\mu}}
       {cr(1-\hat{r}_Q \cdot \vec{\beta}_Q)}\right|_{t' = t_Q'}
     \end{equation}    
       These potentals differ from the LW ones in (5.22) and (5.23) by only an overall factor
        $\gamma_u$. The derivation of the retarded potential just given is essentially the
         same as that of Jackson~\cite{JackLW}. Integrating first over $d^3 x_J$ and then over $t'$ as in (5.14)
        then apparently gives a result quite different to (5.62) where the order of integration is
        reversed. The problem is complicated by the fact that the region of non-zero charge density
       is in motion, and that for the case of a point charge, because of the relation (5.18)
       the integrals over time and spatial position are not independent. The factor
       $1/(1-\hat{r}_Q \cdot \vec{\beta}_Q)$ resulting from the manipulation of the $\delta$-function
        in the time integral of (5.62) has the effect of multiplying this integral by the ratio
         of the times for which the region of non-zero charge density lies on the backward light cone of
         the test charge, when the former is in motion, or at rest. To see this, consider the simple case
      in which $ \vec{\beta}_Q$ is parallel to $\hat{r}_Q$, so that the source charge is moving
      directly towards the test charge, Suppose the the region of non-zero charge density extends
      for the distance $\ell$ parallel to  $\hat{r}_J$ and that the source charge moves towards the test
      charge with with velocity $c \beta_u$. The distance over which the backward light cone of the test
      charge overlaps the moving charge distribution is then $L = \ell/(1-\beta_u)$, so that\footnote{
       For the generalistation of this calculation to the case of motion with $\vec{\beta}_J$ and $\hat{r}_J$
       non-parallel, leading to the last member of (5.69), see, for example Ref.~\cite{SchwartzLW}
       or Ref.~\cite{Griffiths}.}:
      \begin{equation}
  \frac{{\rm light~cone~overlap~in~motion}}{{\rm light~cone~overlap~at~rest}} = \frac{\ell/(1-\beta_u)}{\ell}
     = \frac{1}{1-\hat{r}_Q \cdot \vec{\beta}_Q}    
    \end{equation}
    Thus (5.68) and the LW potentials suggest that the magnitude of the potential of a moving charge
     is just proportional  to the overlap distance (or time) of the distribution with the backward
    light cone of the field point at which the potential is evaluated. This does not however take 
    into account the effective charge density of the moving distribution. The average charge density
    over the light cone overlap region of the charges  in motion is {\it less} in the ratio
      $\ell/L = (1-\beta_u)$ than when the charges are at rest. This exactly cancels the effect of
     the increased overlap distance or time.
    \par It is now clear why the results (5.19) and (5.68) differ. Performing the $t'$ integration
     in the passage from (5.64) to (5.67) has the effect of averaging over the time interval for which
      some elements of the charge distribution lie on the backward light cone of the field point.
    Therefore if $\rho^*$ represents the charge density at some position when the distribution 
    is at rest (given, for example, by (5.16) for a point charge) the time average density $\bar{\rho}$
     corresponding to $J_{\mu}$ in (5.67), at the same position,  when the charge distribution is in motion
     is 
      \begin{equation}
  \bar{\rho} = (1-\hat{r}_Q \cdot \vec{\beta}_Q)\rho^*
  \end{equation}
     so that the result of (5.19) is recovered.  
   That this is the right result, so that (5.19) is correct
      and (5.22), (5.23) and (5.68) are wrong, becomes quite clear on examining closely the derivation
    of the LW potentials given by Feynman in his `Lectures on Physics'~\cite{FeynLW}. Because this is a 
    very transparent derivation, the error which is made, once it is pointed out, is also very
    transparent!

\begin{figure}[htbp]
\begin{center}
\hspace*{-0.5cm}\mbox{
\epsfysize15.0cm\epsffile{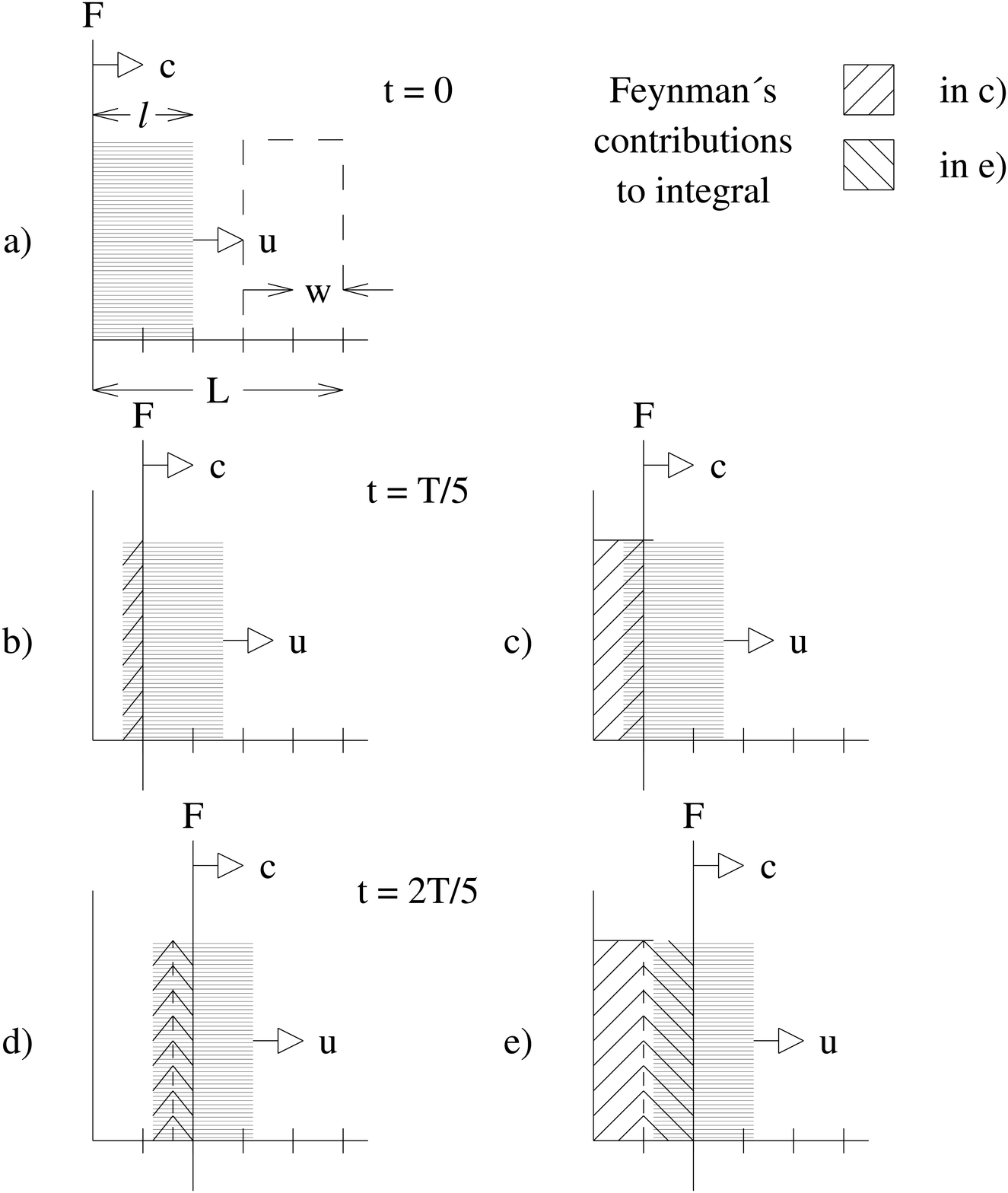}}   
\caption{{\sl Feynman's method of calculating retarded potentials~\cite{FeynLW}.
  A uniform rectangular block of charge of length  {\it l} moves to the right with speed $v$ towards
  a distant field point. The light front, F, in causal connection with the field point, overlaps the block
  for a distance $L$ and a time $T$. In a) F arrives at the front of the block. The position of the block
  when F overtakes it is shown dashed. b) and d) show the positions of F at times $t = T/5$ and $t = 2T/5$
  respectively. The regions of the block sampled by F in the time intervals
  $0 < t < T/5$ and  $T/5 < t < 2T/5$ are shown by the SW-NE and NW-SE cross-hatched areas, respectively.
   The similar crossed-hatched areas in c) and e) show the charge volumes assigned to the potential
    integral, during the same time intervals, in Feynman's calculation. See text for discussion.}}
\label{fig-fig3}
\end{center}
\end{figure}

     \par  Feynman's analysis of the problem of retarded potentials is illustrated in Fig.3.
       A rectangular block of charge of uniform density moves towards the field point, which is sufficiently
     far to the right that the variation of $r_J$ may be neglected in evaluating the integral
     that gives the potential. The light front, $F$, moves across the charge distribution, sampling it.
     Each element of charge which is crossed by $F$ gives a contribution to the potential. The depth of the 
    block of charge is $\ell$ and, as above, $F$ moves over the distance $L$ while crossing the charge 
    distribution. The passage of the front over the charge distribution takes time $T$. Following 
    Feynman, the distance $L$ is divided into bins of width $w$ and the contribution to the potential
    of each bin is considered separately. In Fig.3 the dimensions and velocity $u$ are chosen so that:
     \[ \ell = \frac{2 L}{5},~~~~~w = \frac{L}{5} \]
     It then follows that $u = 3c/5$. In this figure, the positions of the charge distribution
     and the front $F$ at times 0, $T/5$, $2T/5$ respectively are shown. In Figs.3b,3d the front
    has crossed charge thicknesses of $0.4 w$,$0.8w$ respectively. The region crossed during the time
    $0 < t < T/5$ is shown by SW-NE\footnote{The points of the compass: South-West (SW), North-East (NE), 
     North-West (NW) and South-East (SE).}diagonal cross-hatching, that crossed in   $T/5< t < 2T/5$
      by NW-SE diagonal cross-hatching. Thus the average charge density in each bin is reduced, in comparison
    with the situation when the charges are at rest, by 60$\%$. Integrating first over time,
     for each bin, (5.64) then gives:
  \begin{equation}
    \Ai_{\mu} = \frac{u_{\mu} S}{c r_J} \sum_{bins}w \bar{\rho} = \frac{u_{\mu} S L \bar{\rho}}{c r_J}
   \end{equation}
 where $S$ is the surface area of the charge distribution normal to its direction of motion and
   $\bar{\rho}^*$ is the average charge density. From the geometry of Fig.3a,  $\bar{\rho} = 2 \rho^*/5$ where 
   $\rho^*$ is the rest frame charge density. Since $L = 5 \ell /2$ (5.70) gives: 
 \begin{equation}
    \Ai_{\mu} = \frac{u_{\mu} S \ell \rho^*}{c r_J} = \frac{u_{\mu} Q}{c r_J}
   \end{equation}
    where $Q$ is the total charge in the block. Allowing for the propagation time delay of the light front
   with respect to the time of the field point (5.72) agrees with Eqn(5.19) but not with (5.68) or the LW potentials
   in (5.22) and (5.23).
    \par The contributions to the integral given by the first two bins, according to Feynman's original
    calculation~\cite{FeynLW} are shown by the SW-NE and NW-SE diagonal hatching in Figs.3c and 3e respectively.
    The movement of the charge distribution is neglected, and with it the change in the effective
     charge density. Feynman's result is given by replacing $\bar{\rho}$ in (5.71) by $\rho^*$, the density
     of the charge distribution at rest. This gives a result consistent with (5.68), (5.22) and (5.23), but
     is evidently wrong, since charge elements are multiply counted during the passage of the light front.
     For example, a contribution to the integral is assigned proportional to the area of
    the cross-hatched region to the left of $F$ in Fig.3c for $t \le T/5$.
    However, inspection of Fig.3b, showing the actual geometrical
    configuration at $t = T/5$, shows that, because of the parallel
    motion of the charge distribution, $F$ has crossed only the fraction of the region
    in Fig.3c that is both shaded and cross-hatched, not the entire cross-hatched region.
   
 In fact, careful inspection of Fig21-6(c) of Ref.~\cite{FeynLW} shows clearly that
   the contribution due to the passage of the light front over the first bin is overestimated. Only the region
   of the charge distribution to the left of the light front as shown in this figure has been sampled
   at this time, not the filled first bin of  Fig21-6(b).    
   \par A graphic analogue of the sampling of the charge distribution by the moving light front as shown in Fig.3
   was given in Ref.~\cite{PPLW}. The conclusion drawn, however, was just the opposite of the correct one.
   The light front was compared to a group of census takers and the charge distribution to the population 
   of a region. If the census takers move at a certain speed, counting the people they see every day
   then the relative populations of two regions with the same population density are in the ratio of
    the times taken by the census takers to cross the regions. If however, the populations are themselves
    migrating, this procedure will over- (or under-estimate) the true population depending on whether
    the migration of the population is with (or against) the direction of motion of the census takers.
    Instead of identifying (as they should have) the correct sampling of a charge distribution
    with the correct estimation of the population by the census takers, i.e. correcting
    for the effects of migration,  the authors of Ref.~\cite{PPLW}
    claimed instead that integral for the potential should instead be associated with the biased
    population estimates produced by neglect of the migration of the population!
    \par A different and somewhat arcane `derivation' of the LW potentials is presented in Ref.~\cite{LLLW}.
      This employs the usually reliable method of expressing an equation valid in a particular
     reference frame in terms of 4-vectors, in order to obtain a covariant or invariant equation.
   Just this method was used to derive the Lagrangian (1.1) by the present author. It was noted that
    (in the notation of the present paper) the retardation condition (5.18) may be used to write
 the temporal component of $\Ai$, in the frame S$^*$, as :
   \begin{equation}
    \Ai_0 = \frac{Q}{c(t-t_Q')}
  \end{equation}
   It was then noted that the 4-vector:
  \begin{equation}
    \Ai(LW) \equiv \frac{Q u}{x_{ret} \cdot u}
  \end{equation}
 where
  \begin{equation}
    x_{ret} \equiv [c(t-t_Q');\vec{x}_q-\vec{x}_Q(t_Q')]
  \end{equation}
   reduces to (5.73) in the rest frame, S$^*$, of the source charge.
  Eqn(5.74) gives exactly the LW potentials  of (5.22) and (5.23). The overall factor $\gamma_u$ 
 in $u$ cancels from the numerator and denominator of (5.74).  This is certainly true.
  But it is equally true that the solution (5.19) of the d'Alembert equations (5.12)
   and (5.13) also gives the relation (5.73) in S$^*$. How then to choose? The potential
   in (5.19) has been derived from Maxwell equations and the Lorenz condition, whereas
   (5.74) is simply an ansatz; curious, but with no necessary physical relevance. 

 \SECTION{\bf{Summary and Outlook}}
  Intercharge forces and electric and magnetic force fields have been considered in three different
  inertial frames:  S$^*$, where the `source charge' $Q$  is at rest and the `test charge' $q$
  is in uniform motion, S, where both source and test charges are in uniform, parallel, motion,
   and S', where the test charge is a rest and the source charge in uniform motion. 
    \par In Section 1, the equations of motion of two charges undergoing mutual electromagnetic 
     interaction, as derived in Ref.~\cite{JHF1} are recalled. These equations are obtained
     from only Coulomb's law, special relativistic invariance and Hamilton's Principle, without
    invoking any `force field' concept. In 
    accordance with the prediction of QED
  ~\cite{JHF1}, the interaction is assumed to 
   be instantaneous. The equations (1.2) and (1.3) provide a complete relativistic description
   of the motion in terms of the values of the charges, their masses, their velocities, and their
    instantaneous spatial separation. 
   \par The space-time transformation properties, between the frames S$^*$, S and S',
     of the inter-charge forces given by (1.2) and (1.3),
    are discussed in Section 2 and presented in Eqns(2.5)-(2.7).
   \par In Section 3, the transformation properties of electric and magnetic fields derived
     by use of the Lorentz force equation from (1.2) and (1.3),
      referred to as RCED (for Relativistic Classical Electro-Dynamics) fields, are compared to the
     corresponding fields of conventional Classical Electro-Magnetism (CEM). The RCED and CEM force fields
     are derived from the corresponding 4--vector potentials in Appendix A.
       Transformation laws between the frames  S$^*$, S and S'  are 
     presented in (3.13)-(3.15) for the RCED fields and in (3.16)-(3.18) for the CEM fields.
     Similar equations relating transverse (to the direction of motion of the source charge)
     electric and magnetic fields, and transformation laws for magnetic fields, are found in the two cases.
     However, due to the relativistic treatment of the inter-charge separation vector in CEM 
     (in RCED it is Lorentz invariant) no conclusions are possible, at this stage, concerning
     agreement between the transformation equations of longitudinal RCED and CEM  electric fields.
      To help clarify this issue,
     the conventional representation of electric and magnetic fields as elements of a second-rank
     antisymmetric tensor, $F^{\alpha \beta}$, is discussed. It is pointed out that the tensor 
     space-time transformation law of  $F^{\alpha \beta}$ in (3.36)-(3.40) requires
     that each component of $F^{\alpha \beta}$ be a classical field. That is, it must be a function
     of the field point coordinates $x_q$, $y_q$, $z_q$ and the time $t$, each considered to be an independent
      variable. Direct calculation shows agreement between the transformation equations of transverse electric
     fields  and magnetic fields for  $F^{\alpha \beta}$  (assumed to be a `classical field' in the sense explained
     above), in both RCED and CEM. Different results, however, are obtained for the transformation law of the
     longitudinal electric field ($E_x$) in RCED ( $E'_x = (\gamma_u/\gamma_w) E_x$) and for  $F^{\alpha \beta}$ 
     ( $E'_x = E_x$ ). The $E_x$ transformation law for CEM is considered later, after critical
     examination of the standard treatment of relativistic effects in CEM. 
     \par The difference between the the $E_x$ transformation laws of RCED and  $F^{\alpha \beta}$
     is easily understood as a consequence of the fact that neither the 4-vector potential, $A$,
     (of RCED or CEM) nor the electric and magnetic fields obtained from the defining equations (3.19) and (3.20)
      are `classical fields' as far as their dependence on $x_q$ and $t$ is concerned. Due to the dependence of
       $A$ and the fields on the inter-charge separation: $r = |\vec{x}_q-\vec{x}_Q(t)|$, the variables
       $\vec{x}_q$ and $t$ are not independent. In the case that the source charge moves parallel to the
        x-axis, the partial derivatives with respect to $x_q$ and $t$ are related according to Eqn(3.73). Furthermore,
      a discussion of the  intercharge separation vector,
       in the frames S$^*$, S and S, reveals it to be a Lorentz-invariant quantity: $r^* = r = r'$,  $\psi^* = \psi = \psi'$.
       As a consequence, the x-coordinate partial derivatives are also invariant, as shown in Eqn(3.72),
         rather than transforming in a way that mixes spatial and temporal derivatives as in (3.33). Indeed the transformation
     law of (3.33) is valid only when $x_q$ and $t$ are independent variables. This is the case neither for RCED nor CEM.
      Application  of the correct partial derivative relations (3.72) and (3.73) in the defining equations
     (3.19) and (3.20) of the electric and magnetic fields, using the 4-vector transformation properities of $A$ in
     (3.75)-(3.78) verifies all the RCED force-field transformation laws.

     \par Finally, in Section 3, the claimed consistency of the $E_x$ transformation law in (3.18) of CEM with that
      ($E_x' = E_x$) of $F^{\alpha \beta}$ is examined. The relativistic analysis purporting to demonstrate
     the `length contraction' effect (3.87) is shown to be flawed because of the misinterpretation
     of Eqn(3.84) which actually relates a proper time to an apparent time, as in (3.93) or (3.95),
     as a relation connecting two proper times. Correcting the relativistic analysis it is found that the 
     CEM electric field transformation laws (3.96) and (3.97) differ both from those of the classical
     fields  $F^{\alpha \beta}$ and the RCED fields. The correct relativistic analysis of the transformation
     law of the intercharge separation also
     shows that the retarded LW potentials of CEM (unlike the instantaneous RCED ones) do not transform
     as a 4-vector. 
     \par In Section 4 the validity of the Maxwell Equations for the RCED and CEM force fields is investigated
      by direct substitution of the formulae (3.5)-(3.8) for these fields, in the frame S, in the Maxwell Equations.
      As an instructive preamble, hypothetical classical electric and magnetic fields $\vec{\E}$ and $\vec{\B}$
      that, by hypothesis, satisfy the homogeneous free-space Maxwell Equations are first considered.
      The Lorentz covariance of these Maxwell Equations is demonstrated. For the electric field divergence
    equation and the Amp\`{e}re law the covariance is demonstrated by direct Lorentz transformation of 
     the fields and differential space and time operators from the frame S to S'.
       For the magnetic field divergence equation and the Faraday--Lenz 
     law the covariance follows directly from the definitions of electric and magnetic fields, in terms of the 
     4-vector potential, in virtue of 3-vector identities, independently of the space-time functional
    dependence of the potential or the fields. This result holds for both the RCED and CEM force fields. 
     \par Maxwell's derivation of wave equations for the classical fields $\vec{\E}$ and $\vec{\B}$ is
     also recalled. The derivation of the wave equation for the electric field assumes that Gauss' law
      holds for the electric field. Gauss' law for the magnetic field, used to derive the wave equation
     for the magnetic field, is however, as pointed out above, a necessary consequence of the definition,
     (3.20), of the magnetic field and a 3-vector identity, and so does not need to be introduced as a 
     postulate in the derivation of the wave equation.
     \par Direct calculation, presented in Appendix B, shows that while the CEM fields satisfy
       the electric field divergence equation and Amp\`{e}re's law, this is not the case for the RCED
      fields. The covariance of the electric field divergence equation is broken by terms of O($\beta^4$)
      which, as previously pointed out~\cite{JHF2}, lead to a breakdown of Gauss' law for these fields.
      The covariance-breaking term in Amp\`{e}re's law for the RCED fields are of  O($\beta^5$).
       Although the covariance-breaking terms of the electric field divergence equation contain
     contributions from both transverse and longitudinal electric fields, those for Amp\`{e}re's law
      are associated uniquely with the longitudinal electric field. It is demonstrated for both the RCED
     and CEM fields that Amp\`{e}re's law may be derived from the electric field divergence equation,
    given the general relations, (4.20), between transverse electric field components and the magnetic
   field, and the relation (3.73) connecting the $x_q$- and $t$-partial derivatives. Thus in both RCED and CEM
   Amp\`{e}re's law (including Maxwell's added-by-hand `displacement current') is a necessary 
   consequence of the  electric field divergence equation, not an independent law.
    \par Because the CEM fields verify all four Maxwell Equations the derivation of the
     wave equations for electric and magnetic fields are the same as for the hypothetical 
     classical fields $\vec{\E}(\vec{x},t)$ and $\vec{\B}(\vec{x},t)$. Because of the non-covariant
     nature of the RCEM fields however, they do not satisfy wave equations (see Eqns(4.29)-(4.34)).
     \par In Section 5, the force fields discussed in the previous four sections (phenomenological
     manifestatations of the exchange, in QED, of space-like virtual photons between the `source' and
     `test' charges~\cite{JHF1}) are contrasted with radiation fields, that provide a classical 
     description of the QED processes of creation, propagation and destruction of real photons.
      It is suggested that, though the field concept may be avoided completely for the description
     of intercharge forces, it must play an important role in the description of radiation. The standard
     text-book derivation of scalar and vector retarded potentials as solutions of inhomogeneous
      d'Alembert equations, obtained by eliminating either the scalar of the vector potentials from
      the electric field divergence equation (5.10) or Amp\`{e}re's law (5.11) with sources, with the aid
     of the Lorenz condition (5.9), is recalled. Specialising to a point-like source charge yields the 
     retarded potentials of (5.19), differing notably from the retarded Li\'{e}nard-Wiechert potentials (5.22)
     and (5.23) to be found in all test-books on CEM. The reason for this difference (and the incorrectness
     of the LW potentials and the CEM fields derived from them) is explained at the end of Section 5.
    It is due the neglect of the motion-dependence of the time-averaged charge density
    probed by the backward light cone of the field point.
       The retarded potentials of (5.19) are used to derive the retarded electric and magnetic fields
     of (5.20) and (5.21) respectively. These calculations are reported in Appendix C.
    The terms proportional to $1/r^2$ are found to be identical,
        apart from an overall multiplicative factor and a retarded time argument, to the RCEM force
      fields of Eqns(3.5) and (3.6), and quite different to the retarded CEM fields  derived from 
      LW potentials. As is usually done, the terms  proportional to $1/r$ in Eqns(5.20) and (5.21)
      are identified with `radiation fields'. Specialising to a source charge undergoing simple
      harmonic motion, taking the non-relativistic limit, and using the Poynting vector to 
      calculate the time-averaged flux of radiated energy, enables the standard text book
      formula, (5.31), for the energy flux of dipole radiation per unit of solid angle to be derived.
       The close correspondence of this formula in the limit of large distances and low photon
     density to the quantum mechanical path amplitude description~\cite{JHF6} or to photon
     production by an atomic dipole transition~\cite{SakuraiDR} is pointed out. 
      \par The validity of the Lorenz condition is tested by direct substitution
     of the retarded potentials (5.19) in the terms on the left side of (5.9). This calculation
     is also reported in Appendix C. As in the case of the RCED force fields, terms $\propto 1/r^2$ 
      containing the source velocity are found to verify the condition, but not the
      terms $\propto 1/r$, depending on the acceleration of the source charge, which correspond 
      to the radiation fields. In the non-relativistic limit the temporal component of $A$ vanishes
     so that this limit is equivalent to  a gauge where $A_0 = 0$ (Hamiltonian, or temporal gauge~\cite{JackOkun}).
     However this does not
     imply (as is usually assumed) that $\vec{\nabla} \cdot \vec{A} = 0$ (Coulomb gauge). In fact, as shown
     in (5.45),$\vec{\nabla} \cdot \vec{A} = i k A $ where $k$ is the photon wave-number.
      An immediate consequence is that $\vec{A}$ (if interpreted as a polarisation vector)
      is not transverse to the direction of the photon momentum vector. Indeed it is clear
      from inspection of (5.10) that $\vec{A}$ is, instead, parallel to the axis of the oscillating
      dipole source.
.   \par The calculation of the flux of radiated  energy using the Poynting vector is repeated
     in a frame where the dipole source has a relativistic average velocity. The distribution
     is found to be markedly different from that obtained by direct
     relativistic transformation of the distribution from the frame where the mean
     velocity of the oscillating source charge vanishes. The radiation
     fields and the Poynting vector
     therefore only give a correct description of the radiation
     pattern in the non-relativistic approximation, and in this latter frame.

      \par At the end of Section 5, the discrepancy between the retarded 4-vector potential (5.19)
     and the LW potentials (5.22),(5.23) is investigated. It is shown that inverting the order of the
      space and time integrations as in Eqn(5.62) has the effect of weighting the contribution
     of a given spatial point to the retarded potential by a factor proportional to the time of overlap
     of the point in question with the region of non-vanishing charge density.
     The change in the average charge density at the point, due to the motion of the charge, must
      then be taken into account. This has not been done in the calculation of the LW potentials,
     thus overestimating the value of the retarded potential for a  charge distribution moving towards
    the field point at which the potential is to be determined, and underestimating it for a charge
   distribution moving away from the field point. The error is clearly evident on a careful examination of
    Feynman's derivation of the LW potentials~\cite{FeynLW} (see Fig.3). Another `derivation' of the
    incorrect LW potentials~\cite{LLLW} is based on an unsupported relativistic ansatz (5.73) that 
    reduces to the usual Coulomb law (5.72) in the rest frame of the source charge.
 
    \par The RCED formulae derived in Ref.~\cite{JHF1} and Section 5 of the present paper differ in
   many respects from the standard text-book CEM formulae that were proposed or derived in the 19th
   Century before the advent of special relativity theory or quantum mechanics. The sources of these 
  differences are five-fold:
  \begin{itemize}
   \item[(i)] In accordance with the predictions of QED, as discussed in detail in 
               Ref.~\cite{JHF1}, the potential associated with the RCED force fields
            is assumed to be instantanaeous, not retarded according to the speed of
            light in free space, as in CEM.  

 \item[(ii)]  In the standard text-book treatment of the CEM formulae, all originating
              from before the advent of special relativity, relativistic effects, based
              on the space-time Lorentz transformation are misinterpreted. In particular
            a non-existent `length contraction' effect (3.86) is invoked in the discussion of the
            field transformation laws. As shown in previous papers written by the present
            author~\cite{JHFLLT,JHFSSC,JHFCRCS,JHFACOORDS}, such a misinterpretation of the 
            space-time effects of special relativity is
            quite general. All text books on special relativity need to be re-written, 
            as well as the chapters in text books on classical electromagnetism 
            dealing with relativity.
 
 \item[(iii)]  The usual assumption that electric and magnetic fields are `classical'
               ones in the sense that each Cartesian coordinate and the time may be considered
              to be independent variables is manifestly incorrect unless all source charges
              are at rest. The CEM transformation law for the longitudinal electric field:
               $E'_x = E_x$ is therefore incorrect. Unlike magnetic and transverse electric
               fields, the longitudinal electric field is not covariant.  

  \item[(iv)]  There is a simple calculational error (double-counting or under-counting)
                in the derivation of the retarded LW potentials of CEM.

  \item[(v)]  In RCED, unlike CEM, a clear distinction is made between the definitions
                and physical meanings of instantaneous force fields (virtual photon exchange)
               and retarded radiation fields (real photon effects).

     \end{itemize}
\par It is possible that a correct equation is `derived' in a spurious manner, if, for example,
     the effects of two false hypotheses happen to cancel, or if properties are assigned 
   to some hypothetical (and non-existent) physical system result in the derivation 
    of an equation that correctly describes some aspect of another, actually existing, 
    but different, physical system. Indeed, as discussed later, just the second scenario occured in the case of
     Maxwell's `electromagnetic waves'. 
    \par In connection with (i) above, the non-retarded nature of electrodynamical force fields has recently
     been experimentally demonstrated in a conclusive manner~\cite{JAP1}.
     \par The existence of a transverse `electric field in magnetostatics' proportional to $I^2$, as
  predicted by both the  RCED and CEM formulae (3.5) and (3.7), was experimentally demonstrated
    three decades ago~\cite{Edwards}.
    
    \par Concerning (ii), experimental tests are now possible of the `relativity of
          simultaneity' of special relativity, which implies the existence of a
      `length contraction' effect similar to that invoked in the standard relativistic
       interpretation
       of CEM. Two such tests have recently been proposed~\cite{JHFRST}. At the time of
     writing there is no experimental evidence for either relativity of simultaneity, as
    proposed by Einstein, nor the associated length contraction effect. The work presented
   in Refs~\cite{JHFLLT,JHFSSC,JHFCRCS,JHFACOORDS} concludes that both these effects are spurious,
   and do not occur in nature, in contrast to the well-experimentally-verified time
   dilatation effect.
    \par What then is the outlook on the CEM of the late 19th Century, as viewed from the beginning
       of the 21st? Two of the greatest achievements of physics in the 20th Century were special relativity
     theory and quantum mechanics. At mid-century, these two disciplines were combined to create
     Quantum Electro-Dynamics (QED), the most powerful and successful physical theory, within its 
     domain of applicabilty, ever invented by man. In particular, QED provides the constructive basis
    for both Einstein's formulation of special relativity and CEM. Electric and magnetic forces
    result, according to QED, from the exchange of space-like virtual photons. `Electromagnetic waves'
    are, in QED, beams of real photons. From this viewpoint, it was Maxwell, not Einstein who discovered
     the photon. Indeed many essential features of quantum mechanics are are already implicit
   in `electromagnetic waves' interpreted as beams of monochromatic photons~\cite{JHF5}.
   Furthermore, the existence of the photon is crucial for a simple understanding of Einstein's
   development of special relativity. According to special relativistic kinematics, the 
   velocity, $v$, of a particle of rest mass $m$ and momentum $p$ is given by the
    relation:
    \begin{equation}
    v = \frac{p c^2}{(m^2 c^4 + p^2 c^2)^{\frac{1}{2}}}
     \end{equation}
    It follows from this that the velocity of any massless particle, in any inertial frame,
 is the constant $c$. Thus the identification of light with massless particle ---photons,
   renders redundant Einstein's counter-intuitive second postulate of special relativity~\cite{JHF5,JHFLT1}.
    \par However, from a modern viewpoint it can be seen that, fundamentally, special relativity is 
     quite independent of CEM or any other specific dynamical theory, in any domain of physics. This,
    though apparently not generally known, was realised as early as 1910 by Ignatowsky~\cite{Ignatowsky}.
    As shown in Refs~\cite{JHFLT1,JHFLT2}, apart from some weak postulates such as linearity
    or single-valuedness of the space-time transformation equations, either one of the following two postulates
    is sufficient to derive the Lorentz transformation and therefore all the physical consequences
    of special relativity:
     
    \par (I)~~{ \tt Reciprocal space-time measurements of similar measuring rods and \newline clocks in two
             different inertial \newline frames by observers at rest in these frames give identical 
              results}~\cite{JHFLT1}.
    \par (II)~~{ \tt The equations describing the laws of physics are invariant with \newline respect
               to the exchange of space and time coordinates, or, more generally, to the exchange of spatial
               and temporal components of four-vectors}~\cite{JHFLT2}.
   \par In the derivation of the Lorentz transformation based on (I)~\cite{JHFLT1} a parameter, $V$, appears
     naturally in the proof, as the maximum possible relative velocity of two inertial frames. Relativistic
     kinematics, which yields Eqn(6.1), with $c \rightarrow V$,  then shows that V is also the velocity
      of any massless particle in any inertial frame. Hence $c = V$ if light is identified with a massless
      particle. In the derivation based on (II), it is necessary to introduce the parameter, $V$, with the dimensions
    of velocity at the outset in order to define uni-dimensional coordinates to which the space-time exchange
     operation may be applied. Thus $x^0 \equiv Vt$ so that the temporal coordinate  $x^0$ and the spatial
    coordinate $x$ have the same dimension, and may be exchanged.
     \par So, although the Lorentz transformation was discovered as that one which rendered invariant the Maxwell
       Equations in different inertial frames, at the most fundamental level it has no necessary  connection
      with CEM or any other dynamical theory. The role of the Maxwell Equations in the development
      of special relativity is therefore of an essentially heuristic nature. In the same way, from a modern
     viewpoint (i.e. knowing QED), Maxwell's `electromagnetic waves' were a heuristic device pointing towards
     the existence of real photons --massless particles with constant velocity. Actually, as shown above,
    the `electrical'  Maxwell Equations (the electric field divergence law and the electrodynamic Amp\`{e}re law),
     associated with the RCED force fields, are not covariant and there is no associated wave equation. However,
     as previously pointed out~\cite{CSR,JHF1} a clear distinction must be made between `bound' electric and
     magnetic fields, responsible for intercharge forces, corresponding to an instantaneous interaction,
     and constituting a classical, phenomenological, description of the exchange, in QED, of space-like virtual 
     photons, and  retarded `radiation fields' constituting a classical
     phenomenological description of the creation, propagation and destruction of real photons in QED.
      Thus, even if the RCED force fields were covariant, it makes no sense to associate real photons
      with the corresponding `electromagnetic waves'. The subjacent physical process (exchange of space-like
       virtual photons) occurs instantaneously. Such photons cannot be associated with a wave equation 
       predicting propagation at the speed $c$.
     \par In a similar way, `electric field energy' and directed momentum is associated only with the radiation
      fields where it is identical with that transported by real photons. The relativistic kinematics of space-like
    virtual photons shows that they exchange momentum but not energy between interacting charges in their
     overall centre-of-mass frame. However the direction of momentum flow, unlike for that of real photons,
     is ambiguous~\cite{JHF1}. There is therefore no divergent `self energy'  associated with such `bound'
     electric fields. The former arises from the attempt to describe an essentially quantum phenomenon
     (the exchange of virtual photons), at microscopic distance scales, in terms of a classical concept 
      (the electric field) appropriate only as a phenomenological description of the effect of a very large
      number of elementary quantum processes at macroscopic distance scales. 
      \par The non-covariance of the electrical Maxwell Equations for the RCED force fields in (5.1) and (5.4)
   implies the existence of a preferred frame, and therefore a breakdown of the special relativity principle 
   for electromagnetic forces. This lack of covariance applies only to the longitudinal component of electrical
   forces. The `preferred frame' is that in which the source charge is at rest, and as is clear from Eqn(5.4),
 the `covariance-breaking' 4-vector is just the 4-vector velocity of the source charge. 
     \par As demonstrated in Section 5, there is a close connection between the description of real photon
     effects provided by the radiation fields and the quantum mechanical description of radiative 
      processes. This becomes manifest in the `inverse correspondence'~\cite{JHF5} limit of low photon
    density. However, the Poynting vector derived from the radiation fields gives a result consistent
     with the quantum mechanical calculation only in the non-relativistic limit and in a frame where
      the average velocity of the oscillating source charge is much less than the speed of light.
     \par In conclusion, what are the global differences between 21st Century RCED and 19th Century CEM?
        When source charges or current distributions are at rest (i.e. in electrostatics and magnetostatics)
      the two theories are almost indentical. Electric and static magnetic force fields are classical so that
     the usual mathematical treatment (solution of the Laplace Equation) is correct and unchanged. Similarly,
     the magnetic forces between neutral current-carrying conductors are the same, to the lowest order
    in $\beta$, in RCED and CEM. However, in RCED, a neutral current-carrying conductor is a source
    of transverse or longitudinal electric fields $\propto I^2$, as has been experimentally
     demonstrated~\cite{Edwards}. The magnetic field divergence equation and the Faraday-Lenz law
      (both resulting from 3-vector identities) are equally valid in RCED and CEM, as is also the
     Lorentz force equation. However, the electric field divergence equation and the
       electrodynamic Amp\`{e}re law are not obeyed by the RCED force fields, the covariance-breaking
      terms being of O($\beta^4$) and  O($\beta^5$) respectively. In both RCED and CEM the 
      electrodynamic Amp\`{e}re law may be derived from the electric field divergence equation
      and so does not represent an independent physical law.
      In practical (engineering) applications of Maxwell's Equations, where all relativistic corrections
     may be safely neglected, RCED and CEM are completely equivalent.
      \par The radiation fields of RCED and CEM provide a phenomenological description of quantum processes,
       involving real photons, in the limit of very high photon densities. In the non-relativistic limit,
     appropriate to dipole radiation, RCEM and CEM give identical predictions for the radiation pattern.
     All physical effects described by the radiation fields must, however, be ultimately
      derivable from quantum mechanics. As often strongly emphasised by Feynman, according to our best
     current knowledge, the only {\it fundamental} 
        description of the world is that provided by quantum mechanics. Classical physics always 
       corresponds to a definite limit of quantum mechanics. Two examples are $ h \rightarrow 0$ where
      classical mechanics is recovered from quantum mechanics, and the Decoherence limit when the 
     number of interacting elementary quantum systems becomes large and quantum interference effects
     are strongly damped, giving classical statistical mechanics and thermodynamics. The inverse limit, 
     in which it is attempted describe microscopic quantum systems in terms of the macroscopic
     concepts of classical physics is generally not a useful one to consider. The  properties of
     large systems can be derived from those of their elementary constituents, together with
      their mutual interactions of the latter, but the macroscopic classical dynamical variables
       appropriate for the large system are not applicable to the quantum dynamics of a constituent
        microsystem. An example of this is the `self energy' problem for a charged particle of small
        spatial dimensions, where the macroscopically appropriate concept of the electric field
        is applied, in an inappropriate manner, to a microscopic system which must be described, 
        at the fundamental level, by the laws of quantum mechanics. This purely classical divergence problem,
        predating the discovery of quantum mechanics, was inherited by quantum field theory.
       \par In the usual interpretation of CEM, the electric and magnetic fields are assumed to be the 
        fundamental physical entities which `exist' and are observed, whereas the 4--vector potential is only
      a 'non-observable' mathematical construction that may be used to parametrise, in non-unique manner,
       the `physical' fields. This non-uniqueness is encapsulated in the concept of
     `Gauge Invariance'\cite{JackOkun}. On the contrary, in RCED the fundamental physical
     entities are electric charges
      and (as in QED) real and virtual photons. The physical observables are not fields but the intercharge
     forces themselves. Only static classical electric and magnetic force fields may be
      defined, but they are not the fundamental entities of the theory, but rather second-level mathematical
      abstractions, useful to write equations for inter-charge or inter-current forces in a more compact manner.
      The first level mathematical abstraction is the 4--vector potential. But even this is not needed to provide
    a description
     of all mechanical aspects of classical electrodynamics. This is given by Eqns(1.2) and (1.3) which
    contain neither fields nor potentials. The situation is such that the most fundamental objects of the 
    theoretical description (electric charges in motion) are the furthest removed from the observed 
    physical phenomena (the inter-charge forces) that are predicted by the theory.  The 4--vector potential
    is an abstract quantity completely defined by the kinematial and spatial configuration of the
    interacting charges. The force fields are further mathematical abstractions completely defined
    by the  4--vector potential and Eqns(3.19) and (3.20). The observed physical effects,
    the forces acting on the charges, are finally given by the Lorentz force formula in terms of the
    fields. The force fields are therefore, in their definition, very close the the observed 
    physical phenomenon but very far removed from the objects of the fundamental theoretical
   description. The mistake of the creators and present proponents of CEM (as viewed,
    with hindsight and knowledge of QED, from the beginning of the 21st Century) was, following
    Faraday and Maxwell, to assign, in the theory, a fundamental theoretical status to the fields and not to 
    charges and their mutual interaction.
     \par Of course, the inter-charge interaction is completely symmetric between `source' and `test' charges.
    Either one of them can be considered as the `source' of the 4--vector potential and the fields.
    This alone shows that the 4--vector potential and fields do not `exist' in same way that the
    charges do, but are  only useful mathematical abstractions.
     \par The physical meaning of the radiation fields is quite different to that of the force fields. 
      The square of these fields is proportional to the spatial energy density of real photons. These
     fields do not give the force on a test charge. The physical reason for the quadratic dependence
    of the energy density on the field strength is understood by considering the low photon
     density limit at large distances from the source. In this region the radiation fields may be
   identified with the wave function of a real photon~\cite{JHF5}. In can be seen in this way that
    Born's interpretation of the wavefunction in quantum mechanics is already implicit in RCED or CEM
    in the limit of very high photon densities where the square of the radiation field
     measures not the probability to find a photon in the region of a field point, but the energy
    density of photons there.
     \par The work leading to the results presented in the present paper provided many surprises (not to
     say shocks!) for the author. Chief among these was finding the mathematical error in the derivation
    of the LW potentials. Why has it taken more than a century to notice this simple mistake? The message of this paper
    has been that we should use the many advances in our understanding of physics made during the 20th Century
    (special relativity, particle physics and QED) to illuminate and comprehend at a deeper level
    the classical physics of the 19th Century. Hopefully, in the future, such an approach will permeate both
    yet-to-be-written text books and the way that physics is taught to new generations.
 
\newpage
               
 {\bf Appendix A}
\renewcommand{\theequation}{A.\arabic{equation}}
\setcounter{equation}{0}
 \par The RCED and CEM electric and magnetic fields in the frame S are here derived from the defining
  equations (3.19) and (3.20) and the 4-vector potentials: (3.49),(3.50) for the RCED fields, and
  (3.51),(3.52) for the CEM fields. The RCED potentials may be obtained, as shown below, from the electrostatic
  Coulomb potential, whereas the `present time' CEM potentials are derived from the retarded Li\'{e}nard-Wiechert
  ones, as described, for example, in Ref.~\cite{PPLW}.
   In the frame S$^*$, the 4-vector potential has the form $A^* = (Q /r^*;0,0,0)$. Since the intercharge separation is a
    Lorentz scalar (see Section 3 above), $r^* = r$, the potentials in the frame S (3.49) and (3.50), are directly obtained
    by Lorentz transformation of $A^*$  into the frame S, yielding $A = (\gamma_uQ/r;\gamma_u \beta_u/r,0,0)$.
   Substituting this potential into (3.19) gives:
  \begin{equation}
   E_x(RCEM) = -\gamma_u Q\left[ \frac{\partial ~}{\partial x_q}+\frac{\beta_u}{c}\frac{\partial ~}{\partial t} \right]
     \left(\frac{1}{r}\right) =  -\gamma_u Q(1-\beta_u^2) \frac{\partial ~}{\partial x_q} \left(\frac{1}{r}\right) 
      = \frac{Q \cos\psi}{\gamma_u r^2}
  \end{equation}
    where the relation (3.83) that reflects the intrinsically non-classical nature of the force fields and the associated
    potential has been used. Here the geometry of Fig.1, where the vector $\vec{r}$ in the x-y plane, is taken into
    account so that
    \[r^2 = (x_q-x_Q)^2+(y_q-y_Q)^2  \]
    and 
     \[\frac{\partial r}{\partial x_q} = \frac{x_q-x_Q}{r} = \cos \psi,~~~  
       \frac{\partial r}{\partial y_q} = \frac{y_q-y_Q}{r} = \sin \psi \]
     Similarly
\begin{equation}
  E_y(RCEM) = -\gamma_u Q \frac{\partial ~}{\partial y_q}
     \left(\frac{1}{r}\right) = \frac{\gamma_u Q \sin \psi}{r^2}
  \end{equation}
  and
 \begin{equation}
   B_z(RCEM) =  - \frac{\partial A_x}{\partial y_q} =  -\gamma_u \beta_u Q \frac{\partial ~}{\partial y_q}
     \left(\frac{1}{r}\right) = \frac{\gamma_u  \beta_u Q \sin \psi}{r^2}
 \end{equation}
  \par The CEM fields are given by similar calculations on making the replacemants
\begin{equation}
  \gamma_u \rightarrow 1,~~~r \rightarrow r g_u \equiv r(1-\beta_u^2 \sin^2 \psi)^{\frac{1}{2}}
 \end{equation}
    in the above equations.
     In addition, the following derivative is required:
\begin{equation}
  \frac{\partial~ }{\partial X}\left(\frac{1}{r g_u}\right) = -\frac{1}{r^2 g_u}\frac{\partial r }{\partial X}
     -\frac{1}{r g_u^2 }\frac{\partial g_u }{\partial X}
 \end{equation}
   where $X = x_q,y_q$.
  The definition of $g_u$ gives:
    \[\frac{\partial g_u}{\partial X} = -\frac{\beta_u^2 \sin \psi}{g_u} \frac{\partial \sin \psi}{\partial X} \]
    Since
    \[ \frac{\partial \sin \psi}{\partial x_q} = -\frac{\sin \psi \cos \psi}{r},~~~
     \frac{\partial \sin \psi}{\partial y_q} = \frac{\cos^2 \psi}{r} \]
    it follows that
  \[\frac{\partial g_u}{\partial x_q} = \frac{\beta_u^2 \sin^2 \psi \cos \psi}{r g_u},~~~
     \frac{\partial g_u}{\partial y_q} = -\frac{\beta_u^2 \sin \psi \cos^2 \psi}{r g_u} \]
    Substituting these relations into (A.5) gives:
  \begin{eqnarray}
    \frac{\partial~ }{\partial x_q}\left(\frac{1}{r g_u}\right) & = &  -\frac{1}{r^2 g_u}\cos \psi-
     \frac{\beta_u^2 \sin^2 \psi \cos \psi}{r^2 g_u^3} \nonumber \\
           & = & -\frac{ [g_u^2 +\beta_u^2 \sin^2 \psi)] \cos \psi }{r^2 g_u^3} \nonumber \\
           & = & -\frac{\cos \psi}{r^2 g_u^3}
   \end{eqnarray} 
  \begin{eqnarray}
    \frac{\partial~ }{\partial y_q}\left(\frac{1}{r g_u}\right) & = &  -\frac{1}{r^2 g_u}\sin \psi+
     \frac{\beta_u^2 \sin \psi \cos^2 \psi}{r^2 g_u^3} \nonumber \\
           & = & \frac{[-g_u^2 +\beta_u^2(1-\cos^2 \psi)] \sin \psi}{r^2 g_u^3} \nonumber \\
           & = & -\frac{\sin \psi}{\gamma_u^2 r^2 g_u^3}
   \end{eqnarray}
   On making the subsitutions (A.4) into the second members of (A.1), (A.2) and (A.3) and using (A.6) and (A.7), it follows that
   \begin{eqnarray}
    E_x(CEM) & = & \frac{\cos \psi}{r^2 \gamma_u^2 g_u^3}
     = \frac{\cos \psi}{r^2 \gamma_u^2(1-\beta_u^2 \sin^2 \psi)^{\frac{3}{2}}}  \\
    E_y(CEM) & = & \frac{\sin \psi}{r^2 \gamma_u^2 g_u^3}
     = \frac{\sin \psi}{r^2 \gamma_u^2(1-\beta_u^2 \sin^2 \psi)^{\frac{3}{2}}}  \\   
    B_z(CEM) & = & \frac{ \beta_u \sin \psi}{r^2 \gamma_u^2 g_u^3}
     = \frac{ \beta_u \sin \psi}{r^2 \gamma_u^2(1-\beta_u^2 \sin^2 \psi)^{\frac{3}{2}}}
    \end{eqnarray} 
   which are equivalent to Eqns(3.7) and (3.8) of the text.

\newpage
               
 {\bf Appendix B}
\renewcommand{\theequation}{B.\arabic{equation}}
\setcounter{equation}{0}
  \par  Here the validity of the free-space Maxwell equation (4.1) is tested, for the case of RCED and CEM 
   force fields generated by a uniformly moving charge, by direct substitution of the fields in the frame
   S, given by Eqns(3.5)-(3.8), into Eqn(4.1). Since the Maxwell equations involve derivatives of the fields,
   and because the derivative of a vanishing quantity does not necessarily vanish, full account must be taken
   of the 3-dimensional geometry of the fields. This is done by rotating the coordinate systems in (3.2)-(3.12)
    by an angle $\phi$ about the x-axes as in Eqns(4.18) and (4.19). The angle $\phi$ is that between the y-axis
   and the projection of the radius vector connecting the source and test charges onto the y-z plane. 
    The following definitions are used:
   \begin{equation}
    r^2 \equiv  (x_q-x_Q)^2 +(y_q-y_Q)^2 +(z_q-z_Q)^2
   \end{equation}
   \begin{equation}
    r_{yz}^2 \equiv (y_q-y_Q)^2 +(z_q-z_Q)^2
 \end{equation}
  \begin{equation}
  \sin \psi \equiv \frac{r_{yz}}{r},~~~ \cos \psi \equiv \frac{x_q -x_Q}{r},~~~ \sin \phi \equiv \frac{z_q -z_Q}{r_{yz}},
    ~~~\cos \phi \equiv \frac{y_q -y_Q}{r_{yz}}
 \end{equation}
    as well as the partial derivatives:
\begin{equation}
 \frac{\partial r}{\partial x_q} = \cos \psi,~~~ \frac{\partial r}{\partial y_q} = \sin \psi \cos \phi,~~~
 \frac{\partial r}{\partial z_q} = \sin \psi \sin \phi
 \end{equation}
\begin{equation}
  \frac{\partial \sin \psi}{\partial x_q} = -\frac {\sin \psi \cos \psi}{r},~~~
  \frac{\partial \sin \psi }{\partial y_q} = \frac {\cos^2 \psi \cos \phi}{r},~~~
 \frac{\partial \sin \psi }{\partial z_q} = \frac {\cos^2 \psi \sin \phi}{r}
 \end{equation}
\begin{equation}
 \frac{\partial \cos \psi}{\partial x_q} = \frac {\sin^2 \psi}{r},~~~
  \frac{\partial \cos \psi }{\partial y_q} = -\frac { \sin \psi \cos \psi \cos \phi}{r},~~~
 \frac{\partial \cos \psi }{\partial z_q} = -\frac { \sin \psi \cos \psi \sin \phi}{r}
 \end{equation}
\begin{equation}
  \frac{\partial \sin \phi}{\partial x_q} = 0,~~~
  \frac{\partial \sin \phi }{\partial y_q} =-\frac {\cos \phi \sin \phi}{r \sin \psi},~~~
  \frac{\partial \sin \phi }{\partial z_q} = \frac {\cos^2 \phi}{r \sin \psi}
 \end{equation}
\begin{equation}
  \frac{\partial \cos \phi}{\partial x_q} = 0,~~~
  \frac{\partial \cos \phi }{\partial y_q} =\frac { \sin^2 \phi}{r \sin \psi},~~~
  \frac{\partial \cos \phi }{\partial z_q} = -\frac {\cos \phi \sin \phi}{r \sin \psi}
 \end{equation}
  Considering first the RCED fields, and differentiating Eqns(A.1) and (A.2):
  \begin{eqnarray}
  \frac{\partial E_x }{\partial x_q} &  =  & \frac{Q}{\gamma_u}\left( -\frac{2 \cos \psi}{r^3} 
  \frac{\partial r }{\partial x_q} + \frac{1}{r^2} \frac{\partial \cos \psi}{\partial x_q}\right) \nonumber \\
    & = & \frac{Q}{\gamma_u r^3}(-2 \cos^2 \psi + \sin^2 \psi) \\
 \frac{\partial E_y }{\partial y_q} &  =  & Q \gamma_u\left( -\frac{2 \sin \psi \cos \phi}{r^3} 
  \frac{\partial r }{\partial y_q} + \frac{\cos \phi}{r^2} \frac{\partial \sin \psi}{\partial y_q}
   +  \frac{\sin \psi}{r^2} \frac{\partial \cos \phi}{\partial y_q}\right) \nonumber \\
    & = & \frac{Q \gamma_u}{r^3}(-2 \sin^2 \psi \cos^2 \phi + \cos^2 \psi \cos^2 \phi +\sin^2 \phi) \\  
\frac{\partial E_z }{\partial z_q} &  =  & Q \gamma_u \left(-\frac{2 \sin \psi \sin \phi}{r^3} 
  \frac{\partial r }{\partial z_q} + \frac{\sin \phi}{r^2} \frac{\partial \sin \psi}{\partial z_q}
   +  \frac{\sin \psi}{r^2} \frac{\partial \sin \phi}{\partial z_q}\right) \nonumber \\
    & = & \frac{Q \gamma_u}{r^3}(-2 \sin^2 \psi \sin^2 \phi + \cos^2 \psi \sin^2 \phi +\cos^2 \phi)
 \end{eqnarray}    
   Adding (B.9), (B.10) and (B.11) gives:
  \begin{eqnarray}
  \vec{\nabla} \cdot \vec{E}(RCED) & = & \frac{Q}{r^3}\left[\frac{1}{\gamma_u}(-2 \cos^2 \psi+ \sin^2 \psi)
                        + \gamma_u(-2 \sin^2 \psi+\cos^2 \psi +1) \right] \nonumber \\
      & = &  \frac{Q}{r^3}\left(\gamma_u-\frac{1}{\gamma_u}\right)(2-3\sin^2 \psi) 
  \end{eqnarray} 
  \par For the CEM fields the calculation proceeds in a similar fashion with the replacements $\gamma_u \rightarrow 1$,
   $r^2 \rightarrow r^2 \gamma_u^2(1-\beta_u^2 \sin^2 \psi)^{\frac{3}{2}} =  r^2 \gamma_u^2 f_u$ in the RCED equations.
    It follows
   from Eqn(B.12) that the contribution from the r-, $\psi$- and $\phi$-derivatives vanishes in this case.
   The remaining contribution from the $f_u$-derivatives is given by: 
  \begin{eqnarray}
 \left. \frac{\partial E_x }{\partial x_q}\right|_{r,\psi,\phi} &  =  & -\frac{Q \cos \psi}{r^2 \gamma_u^2 f_u^2}
    \frac{\partial f_u }{\partial x_q} \nonumber \\
 &  =  & \frac{ 3 Q \cos \psi}{r^2 \gamma_u^2 f_u^2} \beta_u^2 \sin \psi
    \frac{\partial \sin \psi }{\partial x_q} \nonumber \\
 &  =  & -\frac{ 3 Q  \beta_u^2 \sin^2 \psi \cos^2 \psi}{r^2 \gamma_u^2 f_u^2} \\
  \left. \frac{\partial E_y }{\partial y_q}\right|_{r,\psi,\phi} &  =  & -\frac{Q \sin \psi \cos \phi}{r^2 \gamma_u^2 f_u^2}
    \frac{\partial f_u }{\partial y_q} \nonumber \\
  &  =  & \frac{ 3 Q \sin^2 \psi \cos \phi}{r^2 \gamma_u^2 f_u^2} \beta_u^2
    \frac{\partial \sin \psi }{\partial y_q} \nonumber \\
 &  =  & \frac{ 3 Q  \beta_u^2 \sin^2 \psi \cos^2 \psi \cos^2 \phi}{r^2 \gamma_u^2 f_u^2} \\
  \left. \frac{\partial E_z }{\partial z_q}\right|_{r,\psi,\phi} &  =  & -\frac{Q \sin \psi \sin \phi}{r^2 \gamma_u^2 f_u^2}
    \frac{\partial f_u }{\partial z_q} \nonumber \\
  &  =  & \frac{ 3 Q \sin^2 \psi \sin \phi}{r^2 \gamma_u^2 f_u^2} \beta_u^2
    \frac{\partial \sin \psi }{\partial z_q} \nonumber \\
 &  =  & \frac{ 3 Q  \beta_u^2 \sin^2 \psi \cos^2 \psi \sin^2 \phi}{r^2 \gamma_u^2 f_u^2}
  \end{eqnarray} 
  Adding (B.13), (B.14) and (B.15) gives:
 \begin{equation}
 \left .\left( \frac{\partial E_x }{\partial x_q}+ \frac{\partial E_y }{\partial y_q}+ \frac{\partial E_z }{\partial z_q}
   \right)\right|_{r,\psi,\phi} = \frac{ 3 Q \beta_u^2}{r^2 \gamma_u^2 f_u^2}[-\sin^2 \psi \cos^2 \psi + 
     \sin^2 \psi \cos^2 \psi (\cos^2 \phi + \sin^2 \phi)] = 0
  \end{equation}
  Thus adding (B.12) with $\gamma_u = 1$ to (B.16) gives:
\begin{equation}
  \vec{\nabla} \cdot \vec{E}(CEM) = 0
  \end{equation}
 and so the first Maxwell equation is valid for the CEM force field of a uniformly moving charge.

\newpage
               
 {\bf Appendix C}
\renewcommand{\theequation}{C.\arabic{equation}}
\setcounter{equation}{0}
  \par The geometry of Fig. 1b is assumed with the radius vector $\vec{r}$ in the x-y plane and the source
 charge $Q$ in motion parallel to the x-axis. Thus
\begin{equation}
 r^2 = [x_q-x_Q(t')]^2 + y_q^2
 \end{equation}
  With $\Ai_0$ given by (5.19):
\begin{equation}
   \left. \frac{\partial \Ai_0}{\partial x_q}\right|_{t} = Q\left(-\frac{\gamma_u}{r^2}+\frac{1}{r}
   \left.\frac{\partial \gamma_u}{\partial r}\right|_{t}\right)
    \left. \frac{\partial r}{\partial x_q}\right|_{t}
\end{equation}
  From the definition of $\gamma_u$ before (2.1):
\begin{equation}
  \left.\frac{\partial \gamma_u}{\partial r}\right|_{t} = 
 \beta_u\gamma_u^3 \left.\frac{\partial \beta_u}{\partial r}\right|_{t}
\end{equation}
  It follows from (5.18) that:
\begin{equation}
  \left.\frac{\partial ~}{\partial r}\right|_{t} = -\frac{1}{c}  \left.\frac{\partial ~}{\partial t'}\right|_{t}
\end{equation}
  Combining (C.2)-(C.4) gives:
\begin{equation}
 \left. \frac{\partial \Ai_0}{\partial x_q}\right|_{t} = Q \gamma_u\left(-\frac{1}{r^2}
   -\frac{\beta_u \gamma_u^2 \dot{\beta}_u}{rc} \right)      
 \left. \frac{\partial r}{\partial x_q}\right|_{t}
\end{equation}
   Differentiating (C.1) with respect to $x_q$:
\begin{equation}
 r\left. \frac{\partial r}{\partial x_q}\right|_{t} =  (x_q-x_Q(t'))\left(1-
    \frac{d x_Q(t')}{d t'} \left. \frac{\partial t'}{\partial x_q}\right|_{t}\right)
\end{equation}
  Differentiating (5.18) with respect to $x_q$:
\begin{equation}
 \left.\frac{\partial t'}{\partial x_q}\right|_{t} = -\frac{1}{c}  \left.\frac{\partial r}{\partial x_q}\right|_{t}
\end{equation}
 Combining (C.6) and (C.7) and noting that $d x_Q(t')/dt' = c \beta_u$:
\begin{equation}
  \left. \frac{\partial r}{\partial x_q}\right|_{t} = \frac{x_q-x_Q(t')}{r(1-\hat{r} \cdot \vec{\beta}_u)}
\end{equation}
    Substituting (C.8)  into (C.5) gives:
\begin{equation}
   \left. \frac{\partial \Ai_0}{\partial x_q}\right|_{t} = -\frac{Q \gamma_u}{1-\hat{r} \cdot \vec{\beta}_u}
   \left(\frac{1}{r^3}
   +\frac{\beta_u \gamma_u^2 \dot{\beta}_u}{r^2 c}\right) (x_q-x_Q(t'))
\end{equation}
    Similarly
\begin{equation}
  \left. \frac{\partial \Ai_0}{\partial y_q}\right|_{t} = -\frac{Q \gamma_u}{1-\hat{r} \cdot \vec{\beta}_u}
   \left(\frac{1}{r^3}
   +\frac{\beta_u \gamma_u^2 \dot{\beta}_u}{r^2 c}\right) y_q
\end{equation}
   Because $\Ai_0$ is independent of $z_q$ it follows from (C.9) and (C.10) that
\begin{equation}
  -\vec{\nabla}\Ai_0 =  \frac{Q \gamma_u}{1-\hat{r} \cdot \vec{\beta}_u}
   \left[\frac{1}{r^2}
   +\frac{\beta_u \gamma_u^2 \dot{\beta}_u}{r c}\right] \hat{r}       
 \end{equation}
   Differentiating $\vec{\Ai}$ given by (5.19) with respect to $t$:
\begin{equation}
     \left.\frac{\partial \vec{\Ai}}{\partial t}\right|_{x_q,y_q} = Q \left(-\frac{\gamma_u \vec{\beta}_u}{r^2}
     \left. \frac{\partial r}{\partial t'}\right|_{x_q,y_q}
      +\left[\frac{\vec{\beta_u} \dot{\gamma}_u + \dot{\vec{\beta}_u} \gamma_u}{r}\right]
  \left. \frac{\partial t'}{\partial t}\right|_{x_q,y_q}\right)
  \end{equation}   
 Differentiating (C.1) with respect to $t'$:
\begin{equation}
 \left. r\frac{\partial r}{\partial t'}\right|_{x_q,y_q} =  -c(x_q-x_Q(t'))\beta_u
 \end{equation}
  or 
 \begin{equation}         
 \left. \frac{\partial r}{\partial t'}\right|_{x_q,y_q} =  -c \hat{r} \cdot \vec{\beta}_u
 \end{equation}
  Differentiating (5.18) with respect to t:
 \begin{equation} 
  \left. \frac{\partial t'}{\partial t}\right|_{x_q,y_q} = 1-\frac{1}{c}
    \left. \frac{\partial r}{\partial t'}\right|_{x_q,y_q} \times  \left. \frac{\partial t'}{\partial t}\right|_{x_q,y_q}  
 \end{equation}
  Combining (C.14) and (C.15) and rearranging:
\begin{equation}
 \left. \frac{\partial t'}{\partial t}\right|_{x_q,y_q}
  = \frac{1}{1-\hat{r} \cdot \vec{\beta}_u}
 \end{equation}
 Also 
\begin{equation}
  \dot{\gamma}_u = \gamma_u^3 \beta_u \dot{\beta}_u
 \end{equation}
  Substituting (C.14), (C.16) and (C.17) into (C.12) gives:
\begin{equation}
  \left. \frac{1}{c} \frac{\partial \vec{\Ai}}{\partial t}\right|_{x_q,y_q} = \frac{Q \gamma_u}{1-\hat{r} \cdot \vec{\beta}_u}
  \left[ \frac{\vec{\beta}_u (\hat{r} \cdot \vec{\beta}_u)}{r^2}+ 
   \frac{\gamma_u^2 \vec{\beta}_u \beta_u \dot{\beta}_u + \dot{\vec{\beta}_u}}{r}
  \right]
 \end{equation}
  (C.11) and (C.18), when substituted into (5.5), give Eqn(5.20) of the text.
   \par Since $\Ai_y = \Ai_z = 0$ the curl of $\vec{\Ai}$ is given uniquely by:
\begin{equation}
      \left. \frac{\partial \Ai_x}{\partial y_q}\right|_{t} = Q\left(-\frac{\gamma_u \beta_u}{r^2}
     \left. \frac{\partial r}{\partial y_q}\right|_{t}
      +\frac{1}{r}\left[\beta_u \dot{\gamma}_u + \dot{\beta}_u \gamma_u\right]
      \left. \frac{\partial t'}{\partial y_q }\right|_{t} \right)
 \end{equation}
   Analogously to (C.8):
\begin{equation}
  \left. \frac{\partial r}{\partial y_q}\right|_{t} = \frac{y_q}{r(1-\hat{r} \cdot \vec{\beta}_u)} 
 \end{equation}
     while from (5.18) and (C.20)
\begin{equation}
  \left. \frac{\partial t'}{\partial y_q}\right|_{t} = 
   -\frac{1}{c}  \left. \frac{\partial r}{\partial y_q}\right|_{t}  
     = -\frac{y_q}{cr(1-\hat{r} \cdot \vec{\beta}_u)}
  \end{equation}
   Substituting (C.17), (C.20) and (C.21) in (C.19) gives:
 \[ \left. \frac{\partial \Ai_x}{\partial y_q}\right|_{t} = -\frac{Q \gamma_u}{1-\hat{r} \cdot \vec{\beta}_u}
  \left[ \frac{ \beta_u}{r^3}+  \frac{\gamma_u^2)\dot{\beta}_u}{c r^2}
  \right] y_q \]
   or  
\begin{equation}
  \vec{\Bi} = \hat{k} \Bi_z = (\vec{\nabla} \times \vec{\Ai})_z  = 
      - \hat{k}\left. \frac{\partial \Ai_x}{\partial y_q}\right|_{t} = \frac{Q \gamma_u (\vec{\beta}_u \times \hat{r})}
     {1-\hat{r} \cdot \vec{\beta}_u}
  \left[ \frac{1}{r^2}+ \frac{\gamma_u^2 \dot{\beta}_u}{ c \beta_u r} \right]
  \end{equation}
 which is Eqn(5.21) of the text.
    \par Finally, the derivatives appearing in the Lorenz condition (5.9) are evaluated.
    Differentiating the x-component of $\vec{\Ai}$ with respect to $x_q$ gives, from (5.19):
\begin{equation}
 \left. \frac{\partial \Ai_x}{\partial x_q}\right|_{t} = \frac{Q}{r}
  \left(-\frac{\gamma_u \beta_u}{r}+\beta_u \left. \frac{\partial \gamma_u}{\partial r}\right|_{t}
  + \gamma_u \left.\frac{\partial \beta_u}{\partial r}\right|_{t}\right)
  \left. \frac{\partial r}{\partial x_q}\right|_{t} 
  \end{equation}
  Use of (C.3), (C.4) and (C.8) in (C.23) gives:
\begin{equation}
 \left. \frac{\partial \Ai_x}{\partial x_q}\right|_{t} = -\frac{Q \gamma_u (x_q-x_Q(t') }
   {1-\hat{r} \cdot \vec{\beta}_u}
 \left[\frac{\beta_u}{r^3}+\frac{\gamma_u^2 \dot{\beta}_u}{c r^2}\right]
  \end{equation}
    So that
\begin{equation}
  \vec{\nabla} \cdot \vec{\Ai} =   \left. \frac{\partial \Ai_x}{\partial x_q}\right|_{t}
    = -\frac{Q \gamma_u \hat{r}\cdot \vec{\beta}_u}
   {1-\hat{r} \cdot \vec{\beta}_u}
  \left[\frac{1}{r^2}+ \frac{\gamma_u^2 \dot{\beta}_u}{cr}\right]
  \end{equation}
    Differentiating $\Ai_0$ in (5.19) with respect to $t$ gives:
 \begin{equation}       
  \frac{1}{c} \left. \frac{\partial \Ai_0}{\partial t}\right|_{x_q,y_q} = \frac{Q}{cr}
  \left(-\frac{\gamma_u}{r} \left. \frac{\partial r}{\partial t}\right|_{x_q,y_q} +
 \left.\frac{\partial \gamma_u}{\partial t}\right|_{x_q,y_q} \right)
  \end{equation}
    Combining (C.13), (C.15), (C.16) and (C.26):
\begin{equation}
 \frac{1}{c} \left. \frac{\partial \Ai_0}{\partial t}\right|_{x_q,y_q}
    = \frac{Q \gamma_u}{1-\hat{r} \cdot \vec{\beta}_u}
  \left(\frac{\hat{r} \cdot \vec{\beta}_u}{r^2}
    + \frac{\gamma_u^2 \beta_u \dot{\beta}_u}{cr}\right)
  \end{equation}
   Eqns(C.25) and (C.27) are Eqns(5.40) and (5.41) of the text  
\pagebreak
 
\end{document}